\documentclass[aps,pre,twocolumn,amsmath,amssymb,amsfonts]{revtex4}
\usepackage{epsfig}
\usepackage{graphicx}
\usepackage{dcolumn}
\usepackage{bm}
\usepackage{braket}
\usepackage{color}
\usepackage{comment}
\usepackage{ulem}
\usepackage{bm}
\usepackage{soul, xcolor}
\usepackage{cancel}
\begin{document}
\setstcolor{red}

\newcommand{\blue}[1]{{\color{blue}#1}}
\newcommand{\red}[1]{{\color{red}#1}}
\newcommand{\cyan}[1]{{\color{cyan}#1}}
\newcommand{\magenta}[1]{{\color{magenta}#1}}

\setstcolor{red}
\title{%Cascade reversal and its implications for density and magnetic fields in turbulent disks
Compressible two-dimensional turbulence: cascade reversal and sensitivity to imposed magnetic field
}

\author{Itzhak Fouxon$^{1,3}$}\email{itzhak8@gmail.com}
\author{Alexei G. Kritsuk$^{2}$}\email{akritsuk@ucsd.edu}
\author{Michael Mond$^{1}$}\email{mond@bgu.ac.il}

%\affiliation{$^1$ {\color{red}{you can use now BGU affiliation}}}
\affiliation{$^1$ Department of Mechanical Engineering, Ben-Gurion University of the Negev,
P.O. Box 653, Beer-Sheva 84105, Israel}
\affiliation{$^2$ University of California, San Diego, La Jolla, CA 92093-0424, USA} \affiliation{$^3$ Department of Chemical Engineering, Technion, Haifa 32000, Israel} \begin{abstract}

We study the impact of compressibility on two-dimensional turbulent flows, such as those modelling astrophysical disks. We demonstrate that the direction of
cascade undergoes continuous transition as the Mach number $Ma$ increases, from inverse at $Ma=0$, to direct at $Ma=\infty$. Thus, at $Ma \sim 1$
comparable amounts of energy flow from the pumping scale to large and small scales, in
accord with previous data. For supersonic turbulence with $Ma \gg 1$ the cascade is direct, as in
three dimensions, which results in multifractal density field. For that regime ($Ma \gg 1$) we derive a Kolmogorov-type law for potential forcing and obtain an explicit expression for the third order correlation tensor of the velocity. We further show that all third order structure functions are zero up to first order in the inertial range scales, which is in sharp contrast with incompressible turbulence where the third order structure function, that describes the energy flux associated with the energy cascade is non-zero.  The properties of compressible turbulence have significant implications on the amplification of magnetic fields in conducting fluids. We thus demonstrate that imposing external magnetic field on compressible flows of conducting fluids allows to manipulate the flow producing possibly large changes even at small Mach numbers. Thus Zeldovich's antidynamo theorem, by which at $Ma=0$ the magnetic field is zero in the steady state, must be used with caution. Real flows have finite $Ma$ and, however small it is, for large enough values of $I$, the magnetic flux through the disk, the magnetic field changes the flow appreciably, or rearranges it completely. This renders the limit $Ma \to 0$ singular for non-zero values of $I$.   Of particular interest is the effect of the density multifractality, at $Ma\gg 1$ which is relevant for astrophysical disks. We demonstrate that in that regime, in the presence of non-zero  $I$ the magnetic field energy is enhanced by a large factor as compared to its estimates based on the mean field.  Finally, based on the insights described above, we propose a novel two-dimensional Burgers' turbulence, whose three-dimensional counterpart is used for studies of the large-scale structure of the Universe, as a model for supersonic two-dimensional magnetohydrodynamic flows.  

\end{abstract}
\maketitle

The nature as well as the effects of turbulence in astrophysical disks remain not fully understood despite its significant relevance to the theory of accretion. 
%AK: presumably there are observational measurements of the magnetic field in accretion disks. Hence, overall a pure hydro description would be insufficient. So, be careful about what you actually want to say here. I guess, you want to narrow the problem down to alternatives to MRI wrt reproducing the observed accretion rates?
In this paper we provide theoretical considerations that lead to deeper insight and understanding of the structure of compressible turbulence as well as its impact on the amplification of magnetic fields in thin disks, assuming that they can be modelled as two-dimensional flows. 

The degree of flow compressibility is measured by the Mach number $Ma$, defined here as the ratio of a typical velocity at the forcing scale and speed of sound. In the limit of $Ma\to 0$, two-dimensional modeling has far reaching implications. In that parameter regime the flow is effectively incompressible, the density is constant and magnetic fields, possibly present in conducting fluids, asymptotically decay, according to the famous Zeldovich antidynamo theorem \cite{ll6,antidynamo}. Thus, in the absence of external magnetic fields, the fluid is non-magnetized in the steady state and the planar flow is purely hydrodynamic. Immersing the system at $Ma=0$ in a constant vertical magnetic field, however strong it is, has no impact on the flow. This state is stable -- small, possibly three-dimensional,   magnetic field perturbations decay. Furthermore, perturbations of arbitrary magnitude decay also, provided that they do not excite significant vertical flow, which would turn the flow into three-dimensional. %The above, seemingly, introduces a flaw in modelling astrophysical disks by planar flows, which does not exist in similar modeling of Earth's atmosphere - 

The incompressible limit of $Ma\to 0$ seems therefore inadequate for modelling astrophysical disks by planar flows since, in contrast to the Earth's atmosphere, magnetic fields typically do play important role in astrophysical context. 
%in astrophysical context magnetic field is often relevant. Yet, these conclusions need revisiting, once it is recalled 
These conclusions are further supported by recalling that $Ma$ is finite in real flows, and in particular may reach very large values in astrophysical disks.
%Moreover, in disks, it can be large. 
%might change the situation. 
We show here that compressibility may lead to a dramatic change in the structure and nature of the two-dimensional turbulence.

We first discuss pure hydrodynamics and demonstrate that, considering the flow as a function of the Mach number $Ma$, a strong rearrangement occurs as $Ma$ increases. Thus, at $Ma\ll 1$, the flow is approximately incompressible and has constant density. It exhibits inverse cascade of energy, where the flow creates organized larger scale motions starting from the scale at which the energy is injected. In contrast, at high Mach number the cascade is direct and the flow is characterized by a multifractal (i.e. singular and having non-trivial scaling) density field \cite{kritsuk,frisch,fm}. At $Ma=0$ as well as at $Ma=\infty$ the velocity decouples from the density. 

For conducting fluids study of hydrodynamics might be insufficient, as magnetic fields could (spontaneously) arise. Two types of two-dimensional configurations have been studied extensively over the last few decades, mainly numerically. One configuration is characterised by a net magnetic flux that pierces the disk vertically while in the other the net flux is zero 
\cite{Hawley,cattaneo,Lesur,Latter,pessah,Cui,Datta}. Large-scale net magnetic flux in the former case may be provided by the dipole magnetic field of a central accreting object or by its companion, or may be advected into the disk from the interstellar medium. Large-scale magnetic fields provide key ingredients in the generation of outflows that play an important role on the formation of stars and their protostellar disks \cite{cao,Ray}. The zero net flux studies on the other hand aim at exploring the possibility of a self-magnetisation of the disk where such properties as angular momentum transfer do not depend on external magnetic fields. 
%With regard to the possible existence of 

In order to gain a deeper understanding of the role played by the magnetic flux we establish a generalization of Zeldovich's theorem to include the compressibility of planar flows.
%, 
We show that if the magnetic field threading the disk, as measured by the flux $I$, vanishes, 
%(we remark that the flux is a key ingredient in the jet formation model \cite{cao,Ray}). We demonstrate that under the condition $I=0$, 
the 
%AK maybe "fluctuations of magnetic field decay" because we just assumed that there is no mean field, hence what remains is fluctuations.
magnetic field decays at any $Ma$, so that the fluid is non-magnetized in the steady state. However, if the system is immersed in a constant (large-scale) magnetic field, so that 
$I\neq 0$, the situation can be very different. 
In that case a steady-state inhomogeneous non-zero magnetic field 
%AK: What exactly is generated? We consider a magnetized case with an applied non-zero mean B-field. Does this mean field become stronger? Or maybe fluctuations of the magnetic field get amplified? Or maybe both? Please be explicit.
is generated, whose impact on the flow is determined by the magnitude of $I$. We further show that for any $Ma$, there is a range of $I$ for which the back reaction of the magnetic field on the flow is non-negligible. For instance, Zeldovich's result does not apply for arbitrarily small $Ma$, if the net flux $I$ is large enough  obeying $I\gg 1/Ma$ (see refinements below). In that case even though the Mach number is arbitrarily small, the flow may be strongly affected by the magnetic field due to the high values of the magnetic flux. In the opposite limit of large $Ma$, we show that the magnetic energy density is larger than the square of the mean-field by a power of $Ma$ for those $I$ that cause negligible Lorentz forces on the velocity. 

 The above insights may turn useful in progressing toward an understanding of accretion. They also suggest possible terrestrial applications where $Ma$ is typically not that large. Thus for a conducting fluid whose flow can be considered as planar, changing the external magnetic field changes the flow. These changes may include amounts of energy directed to large and small scales. The ease with which external magnetic fields can be controlled 
 %AK English is not clear in this sentence. It sounds like something regulates the external field, while I think the field actually regulates processes in such systems. So, maybe "the ease with which the external field can be manipulated makes ... attractive"?
 makes the study of such prospective of flow manipulation highly attractive.

We use for our study the simplest model that 
%can be hoped to 
captures the relevant physics. The flow is modelled as two-dimensional isothermal compressible turbulence. Generally, two-dimensional fluid mechanics is expected to describe motions on scales larger than the system (disk) 
%AK: width or thickness?
thickness. For those scales, it seems plausible that introduction of three-dimensional perturbations would not change the main features uncovered by the two-dimensional study. This was shown to be true at small Mach numbers, see \cite{guo} and references therein. Furthermore, checking the validity of the isothermal approximation requires first a thorough scrutiny of its consequences and implications. A step towards achieving that goal in the highly compressible turbulence regime is proposed in the current work. 

We start by assuming a pure hydrodynamic framework where there is no magnetic field. We study how the properties of two-dimensional isothermal compressible turbulence change as the Mach number increases from zero to infinity (the Reynolds number is assumed to be large and fixed). We observe that the transfer of kinetic energy changes direction from inverse to direct cascade. This extends the observation of \cite{fk} that as the Mach number increases,  growing portions of the kinetic energy, which at $Ma\ll 1$ is transferred predominantly from small to large scales (inverse cascade), are transferred from large to small scales. Thus at $Ma\sim 1$ the amount of energy transferred down and up the energy injection scale are comparable \cite{fk}. We provide arguments that this trend continues and at $Ma=\infty$ all kinetic energy is transferred down scales, which implies deep changes in the properties of the  velocity and density fields. Our results suggest that, in contrast to the $Ma\ll 1$ case, supersonic turbulent flows in two and three dimensions are rather similar in some respects.

The self-consistency of pure hydrodynamic framework for turbulent flows of conducting fluids needs to be examined. This is due to the possible instability, known as dynamo effect, by which small fluctuations of magnetic fields that are ubiquitous in the Universe, are amplified by the flow \cite{ll8}. As already said, Zeldovich's theorem suggests that neglecting the magnetic field is self-consistent 
%at $Ma\ll 1$, 
in the limit $Ma \to 0$ where the flow could be thought to good approximation to be incompressible. In the opposite limit of $Ma\gg 1$, similarity of supersonic turbulence in two and three dimensional situations raises the possibility that the flow has dynamo action as its three-dimensional counterpart, i.e., amplifies small fluctuations of magnetic field. Moreover, due to compressibility, not only fluctuations but also the mean magnetic field can be relevant. This calls for studying the role of the magnetic field which is done after exploring the pure hydrodynamics case.

% {\bf Fundamental model and its properties}---We begin our considerations from studying the leading order approximation to the turbulent flow at large Mach numbers. Our main purpose is to show that density becomes a multifractal field in this limit. Later this would be shown to have non-trivial implications for the structure of magnetic field in the disk.

{\bf The fundamental model and its properties}---Turbulence is assumed to be driven by external acceleration fields which do not depend on the flow (the external volume force is proportional to the density) and have characteristic scale of spatial variations $L$, the pumping scale. We use the simplest model which is isothermal turbulence with simplified viscous dissipation term. There the pressure $p$ is given by $\rho c_s^2$ where $\rho$ and $c_s$ are the mass density of the disk and the constant speed of sound, respectively. We use dimensionless variables where the velocity is rescaled by its characteristic value $u$ at the scale $L$, the density is rescaled by its mean, time and space coordinates are rescaled by $L/u$ and $L$, respectively. The flow is governed therefore by the following equations
\begin{eqnarray}&&\!\!\!\!\!\!\!\!\!\!\!\!\!\!
\partial_t \rho+\nabla\cdot(\rho\bm v)=0, \label{nsp}\\&&\!\!\!\!\!\!\!\!\!\!\!\!\!\!
\partial_t\bm v+(\bm v\cdot\nabla)\bm v=-\frac{1}{Ma^2}\nabla \ln \rho+\bm a+\nu\nabla^2\bm v,\nonumber
%+\frac{\nabla_k\sigma'_{ik}}{\rho},\nonumber
\end{eqnarray}
where $Ma\equiv u/c_s$, the dimensionless density is $\rho$, and $\bm v\equiv \bm u/u$ with $\bm u$ the dimensional flow. Here $\bm a$ is the dimensionless acceleration and we denote the inverse Reynolds number by $\nu$, which is assumed to be constant. We use a simplified form of the viscosity term since we are interested in the inertial range properties of turbulence where its precise form is believed to be irrelevant \cite{frisch}. We stress that our definition of the Mach number uses velocity at $L$ and not the actual velocity as, e.g., in \cite{fk}. The definitions are similar at $Ma\gtrsim 1$, however differ at $Ma\ll 1$ where, due to inverse cascade, typical velocity is much larger than the velocity at the pumping scale.  

Eqs.~(\ref{nsp}) can be considered as a fundamental model of compressible turbulence. In spite of including compressibility in the simplest possible way, they are widely believed to have significant applications, see e.g. \cite{kritsuk} and references therein. 

In the two-dimensional case the vorticity $\omega$ (given by the only non-vanishing, vertical, component of the velocity curl) obeys the advection-diffusion equation with forcing \begin{eqnarray}&&\!\!\!\!\!\!\!\!\!\!\!
\partial_t\omega+\nabla\cdot (\omega\bm v)=\nu \nabla^2\omega+s. \label{fp}
\end{eqnarray}
Here $s$ is the source of vorticity, given by the only non-vanishing, vertical, component of $\nabla\times\bm a$. The absence in Eq.~(\ref{fp}) of the vorticity stretching term, present in the three-dimensional case, has far-reaching consequences.

{\bf Potential forcing}---One of the consequences of Eq.~(\ref{fp}) is significant simplifications in the case of potential forcing, such for instance as gravity where $\bm a=-\nabla\psi$ with a certain $\psi$. We have then $s=0$ in Eq.~(\ref{fp}) and the equation is of the Fokker-Planck form. If the velocity were independent of $\omega$ then, due to dissipative nature of the Fokker-Planck equations, the steady state would have $\omega\equiv 0$, see \cite{fp}. However, vorticity is a so-called active scalar that reacts on the flow that advects it. This could cause a difference. 
A well-known example is vorticity of three-dimensional turbulence and small fluctuations of the magnetic field. Both obey the same equation of first order in time, however their behavior is very different \cite{review}. In our case however, the existence of entropy functional guarantees the asymptotic decay of $\omega$, that has zero average, see detailed discussion in Ch. III of \cite{fm} and references therein, and also the study of vertical magnetic field below.

Relaxation of $\omega$ to zero signifies that mathematically well-defined potential solutions  of  Eqs.~(\ref{nsp}), with $\bm v=\nabla\phi$, given by gradient of certain potential $\phi$, are stable.  
Thus for potential forcing the flow is going to be potential,
which is not necessarily so in the three-dimensional case where instability is possible. This allows us to rewrite the momentum equation as (cf. with Bernoulli's equation)
\begin{eqnarray}&&\!\!\!\!\!\!\!\!\!\!\!\!\!\!
\partial_t\bm v=-\nabla \left(
\frac{\ln \rho}{Ma^2}+\frac{v^2}{2}+\psi\right)+\nu\nabla^2\bm v,\label{scalingp}
\end{eqnarray}
so that the flow potential $\phi$ obeys 
\begin{eqnarray}&&\!\!\!\!\!\!\!\!\!\!\!\!\!\!
\partial_t\phi+\frac{(\nabla\phi)^2}{2}=-\frac{\ln \rho}{Ma^2}-\psi+\nu\nabla^2\phi. \label{cf}
\end{eqnarray}
Thus, the evolution reduces to two scalar fields, $\rho$ and $\phi$. 

{\bf Scaling relation}---We observe that Eq.~(\ref{scalingp}) has the form of a local conservation law. It can be written as $\partial_t v_i+\nabla\cdot \bm j^i=0$
where the flux functions are given by $(j^i)_k=\delta_{ik}\left(Ma^{-2}\ln\rho+v^2/2+\psi\right)+\nu \nabla_kv_i$. In the free (forceless) non-dissipative case the latter are given by $(j^i)_k=\delta_{ik}\left(Ma^{-2}\ln\rho+v^2/2\right)$, and give rise to
 the following general scaling relation in steady state \cite{us}:
\begin{eqnarray}&&\!\!\!\!\!\!\!\!\!\!\!\!\!\!
\left\langle v_i(0)\left(\frac{\ln \rho(\bm r)}{Ma^2}+\frac{v^2(\bm r)}{2}\right)\right\rangle=\frac{\langle \bm v(0)\cdot\bm a(0)\rangle \bm r}{2}. \label{scaling}
\end{eqnarray}
This equation is obtained by using Eq.~(\ref{scalingp}) in the stationarity condition $\partial_t \langle v_i(0)v_i(\bm r)\rangle=0$. Eq.~(\ref{scaling}) holds in the inertial range of scales, which are below $L$, and yet not too small so that the viscosity is negligible (here and below, for transparency we often use dimensional quantities in qualitative considerations). In the case of inverse cascade (see below) a sustainable steady state may require the presence of a linear dissipation term in the momentum equation, see e.g. \cite{buhler} and cf. \cite{fk} where it is shown that steady state can also hold for Eqs.~(\ref{nsp}) as they are.  Here we refer to cascade in the sense of the transfer of energy. Statistical isotropy is also assumed above, for a form that does not involve the isotropy assumption see \cite{us1}. 

We observe that by spatial homogeneity, which is assumed in our study, we have $\langle \bm v(0)\cdot\bm a(0)\rangle=\langle  \psi(0)\nabla\cdot\bm v(0)\rangle$ in the RHS of Eq.~(\ref{scaling}) where we use $\bm a=-\nabla\psi$. Here arbitrary additive constant in the potential is irrelevant due to $\langle  \nabla\cdot\bm v(0)\rangle=0$. Eq.~(\ref{scaling}) can be rewritten in a standard way as a relation for the difference \cite{frisch}, which will be subject of future investigations. In the limit $Ma\to 0$ it is equivalent to two-dimensional version of the Kolmogorov relation and hence it can be considered as its finite Mach number generalization \cite{us,review}. 

We return to Eqs.~(\ref{nsp}) for generic, not necessarily potential $\bm a$, and study the behavior of turbulence that they describe, as a function of $Ma$,  assuming that all other parameters in the equations are fixed. For this we consider the two opposite limits, of $Ma\to 0$ and $Ma\to\infty$.

{\bf Low Mach number limit}---In the limit of $Ma\to 0$ Eqs.~(\ref{nsp}) reduce to the well-studied equations of incompressible turbulence (there are some delicate points in the case of potential forcing that does not inject solenoidal component of velocity. These are irrelevant for the discussion here), see e.g. \cite{ll6}. 
A fundamental property of incompressible turbulence is that non-linear interactions organize so as to provide a constant flux of energy
$\int v^2d\bm x/2$, see e.g. \cite{frisch}. Thus in the inertial range, defined as the range where both the forcing and the viscosity can be neglected, the energy is transferred between the Fourier modes of the flow at constant total rate. It can flow either from large to small scales (direct cascade) or from small to large scales (inverse cascade) \cite{review}. The direction of the cascade depends on the number of quantities that have a positive density in Fourier space and are conserved by the Navier-Stokes equations without forcing and dissipation terms. Three-dimensional turbulent flow has only one such quantity, which is energy. In this case the quadratic non-linearity of the equations gives rise to a direct energy cascade. In contrast, two-dimensional incompressible turbulence, that holds in the $Ma\to 0$  limit, has two such quantities - energy and enstrophy (integral of squared vorticity). As a result the energy flows from smaller to larger scales, from the injection scale to larger scales, corresponding to the inverse cascade of energy, see e.g. \cite{frisch}. This direction of kinetic energy flow can be understood as arising from interactions of planar vortices and their self-organization, see e.g. \cite{review}.

{\bf Infinite Mach number limit}---We turn now to the opposite limit of supersonic turbulence, $Ma\to\infty$, and demonstrate that the energy cascade is direct there, namely from large to small scales. 

%the viscous stress tensor $\sigma'_{ik}$ is
%\begin{eqnarray}&&\!\!\!\!\!\!\!\!\!\!\!
%\sigma'_{ik}=\nu\rho\left(\nabla_kv_i+\nabla_iv_k-\frac{2}{3}%\delta_{ik}\nabla\cdot\bm v\right)+\zeta\delta_{ik}. %\label{stress}
%\end{eqnarray}
%where $\nu$ is the kinematic viscosity and $\zeta$ is the bulk viscosity. The detailed form of $\sigma'_{ik}$ is irrelevant below. We will model

We observe that if all parameters besides $Ma$ are fixed, and $Ma\to\infty$, the pressure term in Eqs.~(\ref{nsp}) becomes negligible at any given scale $l$ that satisfies $(l Ma^2)^{-1} \ll 1$. Thus, for large enough $Ma$, pressure is negligible at the pumping scale $L$. Large scales (here $\lesssim L$) are then described by the equations
\begin{eqnarray}&&\!\!\!\!\!\!\!\!\!\!\!\!\!\!
\partial_t \rho+\nabla\cdot(\rho\bm v)=0, \label{ns}\\&&\!\!\!\!\!\!\!\!\!\!\!\!\!\!
\partial_t\bm v+(\bm v\cdot\nabla)\bm v=\bm a+\nu\nabla^2\bm v,\nonumber
\end{eqnarray}
obtained by setting $Ma=\infty$ in Eqs.~(\ref{nsp}). In the case of $\bm a=-\nabla\psi$ the flow, as explained above, is potential and the above equations reduce to the so-called Burgers' turbulence \cite{review,Burgulence}. 

The pressure is negligible at large scales because, for supersonic turbulence, inertial degrees of freedom of large scale eddies are much faster than the pressure degrees of freedom, propagating at the speed of sound. Nonetheless, since smaller scale eddies have smaller velocities, then at scales where the flow velocity is of order of the speed of sound or smaller, the pressure is relevant again. Thus, Eqs.~(\ref{ns}) apply only at scales larger than the so called sonic scale $l_s$, defined as the scale where the eddy velocity and the sound velocity are equal. For quantitative definition of $l_s$ the second order structure function of the flow could be used \cite{sonic}. 

{\bf Density is smooth below the sonic scale}---By definition the eddies below $l_s$ are subsonic, so that they act on the density as effectively incompressible velocity. Thus, the density field below $l_s$ is smooth at any Mach number \cite{sonic}. For $Ma\ll 1$ the Kolmogorov scaling of the inverse cascade \cite{guo} gives $l_s\sim L/Ma^3\gg L$ so the density is smooth up to scales much larger than $L$. 

\begin{figure}[ht]
 \centerline{\includegraphics[scale=0.32]{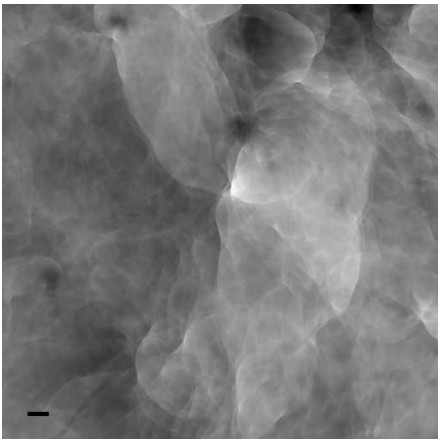}}% Images in 100% size
 \caption{A sample snapshot of the steady state density field from a $4096^2$ simulation of near-isothermal ($\gamma=1.001$) turbulence at 
  $Ma\sim0.1$ (case E from \cite{fk}). The grey-scale ramp follows $\ln\rho$, showing lower densities with darker shades of grey, $\rho\in[0.58,1.60]$. A black scale bar in the lower left indicates the forcing scale. Density varies mostly at larger scales.}\label{Fig:M06}
\end{figure}

\begin{figure}[ht]
 \centerline{\includegraphics[scale=0.32]{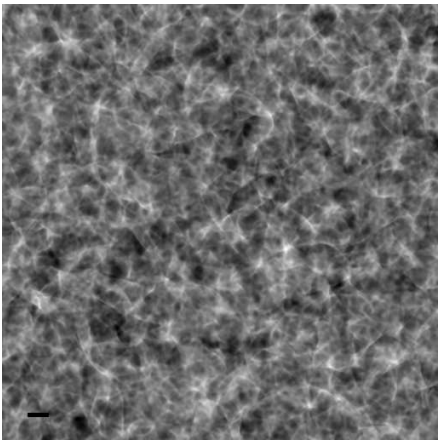}}% Images in 100% size
 \caption{Same as in Fig.~\ref{Fig:M06}, but from a $2048^2$ simulation early in the evolution --- before the system reaches a steady state with $Ma\sim0.1$; $\rho\in[0.94,1.12]$.}\label{Fig:M01}
\end{figure}

Kolmogorov scaling can only be used if eddies at scales larger than $L$ are really subsonic, which, e.g., is not the case of \cite{fk}, where $k^{-2}$ spectrum fits the data better than the Kolmogorov scaling, implying $l_s\sim L/Ma^2$. The data for the case of \cite{fk} is shown in Fig. \ref{Fig:M06}. To determine the Mach number in the Figure, we observed that for parameters of \cite{fk} we have $l_s\sim L 10^{1.7}$ which results in $Ma\simeq \sqrt{L/l_s}\sim 0.1$
%AK: It should be easy to calculate Ma in your terms for case E using the standard formula epsilon_f=u^3/2L, where u is typical velocity at the injection scale L, epsilon_f energy injection rate. We have from Table 1 in supplemental material for [17] epsilon_f=0.008, L=0.012, therefore Ma=0.058 -- which is close to your estimate Ma=0.1. I hope this makes sense.
(we remark that within the definition of the Mach number in \cite{fk} this case corresponded to 
%AK: Table 1 in sullpemental material for [17] gives Ma=0.62 for case E.
$Ma=0.62$). Fig. \ref{Fig:M06} demonstrates that the density field is smooth over scales much larger than the pumping scale. The actual scale of variations is the sonic scale that in this case coincided with the system size. We also provide in Fig. \ref{Fig:M01} the transient state before the steady state is reached. That shows much smaller scale structures that however disappear in the steady state.

\begin{figure}[ht]
 \centerline{\includegraphics[scale=0.32]{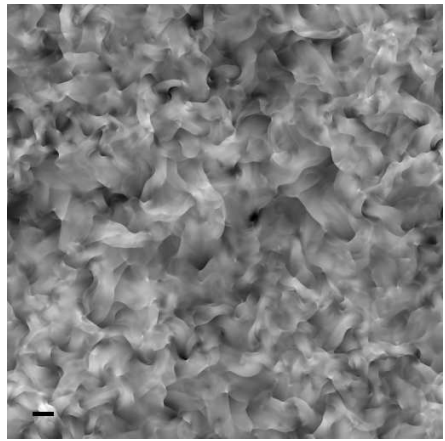}}% Images in 100% size
 \caption{Same as in Fig.~\ref{Fig:M06}, but from a $512^2$ simulation in a statistically steady state at $Ma=1.0$; $\rho\in[0.20,5.11]$. It is seen that density varies over the pumping scale. }\label{Fig:M096}
\end{figure}

In the opposite limit of $Ma\gg 1$, using $k^{-2}$ spectrum of the Burgers' turbulence (see below and \cite{fk,boldyrev,grs}) we find $l_s\sim L/Ma^2\ll L$. This agrees with observations:  at $Ma\simeq 4$ we have $l_s/L\simeq 1/16$ for simulation parameters in \cite{sonic}. The range of scales $l_s\ll r\ll L$, that exists at $Ma\gg 1$, was named in \cite{fm} supersonic inertial range. Finally, by definition $l_s\sim L$ at $Ma\sim 1$, see Fig.~\ref{Fig:M096} (here our definition of $Ma$ and that of \cite{fk} coincide).

{\bf An approach to compressible turbulence}---Similarity of two-dimensional compressible turbulence with "small pressure", i.e. Eqs.~(\ref{nsp}) at $Ma\gg 1$, and Burgers turbulence was mentioned previously in \cite{boldyrev}. It is most remarkable that both at $Ma=0$ and $Ma=\infty$ the velocity decouples from the density and obeys closed equation. Correspondingly, the density becomes a passively advected field. The decoupling allows to study first the properties of velocity separately from the density, and then study the density as in a passive scalar problem \cite{fm,review}. Since velocity-density coupling is a main difficulty introduced by the compressibility then this opens the possibility to describe finite $Ma$ cases by interpolation between the two limits. 

{\bf Kolmogorov-type law for potential pumping}---In the potential case, similarly to Eq.~(\ref{scalingp}) we can write the last of Eqs.~(\ref{ns}) as 
\begin{eqnarray}&&\!\!\!\!\!\!\!\!\!\!\!\!\!\
\partial_t\bm v+\nabla\frac{v^2}{2}=\bm a+\nu\nabla^2\bm v.
\end{eqnarray}
The potential equation becomes
\begin{eqnarray}&&\!\!\!\!\!\!\!\!\!\!\!\!\!\!
\partial_t\phi+\frac{(\nabla\phi)^2}{2}=-\psi+\nu\nabla^2\phi, \label{cf1}
\end{eqnarray} cf. Eq.~(\ref{cf}). 
This equation implies the following evolution law for the pair-correlation function: 
\begin{eqnarray}&&\!\!\!\!\!\!\!
\partial_t\langle \phi(0)\phi(\bm r)\rangle\!+\!\frac{\left\langle \phi(\bm r)v^2(0)\right\rangle\!+\!\left\langle \phi(0) v^2(\bm r)\right\rangle}{2}\!=\!-\langle \psi(0)\phi(\bm r)\rangle
\nonumber\\&&\!\!\!\!\!\!\!
-\langle \phi(0)\psi(\bm r)\rangle
\!+\!\nu \langle \phi(\bm r)\nabla^2 \phi(0)\rangle
%\nonumber\\&&\!\!\!\!\!\!\!\!\!\!\!\!\!
+\nu \langle \phi(0)\nabla^2 \phi(\bm r)\rangle, \label{stationarit}
\end{eqnarray}
%This equation implies the evolution law for the pair-correlation function 
%\begin{eqnarray}&&\!\!\!\!\!\!\!\!\!\!\!\!\!
%\partial_t\langle v_i(0)v_k(\bm r)\rangle\!+\!\left\langle v_k(\bm r)\nabla_i\frac{v^2}{2}(0)\right\rangle\!+\!\left\langle v_i(0) \nabla_k\frac{v^2}{2}(\bm r)\right\rangle
%\nonumber\\&&\!\!\!\!\!\!\!\!\!\!\!\!\!
%=\!\langle a_i(0)v_k(\bm r)\rangle\!+\!\langle v_i(0)a_k(\bm r)\rangle
%\!+\!\nu \langle v_k(\bm r)\nabla^2 v_i(0)\rangle
%\nonumber\\&&\!\!\!\!\!\!\!\!\!\!\!\!\!
%+\nu \langle v_i(0)\nabla^2 v_k(\bm r)\rangle, \label{stationarity}
%\end{eqnarray}
which is valid in the supersonic range of scales, $r\gg l_s$, where the pressure is negligible. We assume that the viscosity is so small that it is also negligible in this range, as discussed previously. We consider the above equation in the steady state, where the time derivative term vanishes. We have, assuming spatial homogeneity and isotropy, that at $r\ll L$
\begin{eqnarray}&&\!\!\!\!\!\!\!\!\!\!\!\!\!
\nabla^2\langle \phi(0)\psi(\bm r)\rangle=
\langle \bm v(0)\cdot \bm a(\bm r)\rangle
\!\approx\! \langle \bm v(0)\cdot \bm a(0)\rangle\!=\!\epsilon,
\end{eqnarray}
%There
%\begin{eqnarray}&&\!\!\!\!\!\!\!\!\!\!\!\!\!
%\langle v_i(0)a_k(\bm r)\rangle\!\approx\! -\langle %v_i(0)\nabla_k\psi(0)\rangle\!=\!\langle \psi(0)\nabla_kv_i(0)\rangle\!=\!\frac{\epsilon\delta_{ik}}{2},
%\end{eqnarray}
where we introduced the pumping power $\epsilon\equiv\langle \bm v(0)\cdot\bm a(0)\rangle$, and used $\bm a=-\nabla\psi$ and $a(\bm r)\approx a(0)$ at $r\ll L$. 
%We also observed that isotropy and parity imply $\langle \psi(0)\nabla_kv_i(0)\rangle=\delta_{ik}\langle \psi(0)\nabla\cdot \bm v(0)\rangle/2$. 
We conclude that in isotropic turbulence Eq.~(\ref{stationarit}) yields in the supersonic inertial range, $l_s\ll r\ll L$, the following lowest order relation:
\begin{eqnarray}&&\!\!\!\!\!\!\!\!\!\!\!\!\!
\left\langle v^2(0)\phi(\bm r)\right\rangle\!=\!-\langle \psi(0)\phi(0)\rangle-\frac{\epsilon r^2}{2d}. \label{ts}
\end{eqnarray}
 We introduced the space dimension $d$ as a useful parameter, in our case $d=2$ (if $d=3$, the consideration serves merely a reference purpose, since potential solutions are probably unstable). The above relation holds at any fixed $r$, which is much smaller than $L$, in the limit of infinite Mach and Reynolds numbers. 

We derive now a Kolmogorov-type relation from Eq.~(\ref{ts}). For that purpose we observe that the three-point correlation function of the velocity potential $\phi$ obeys
\begin{eqnarray}&&\!\!\!\!\!\!\!\!\!
\left\langle \phi(\bm r_1)\phi(\bm r_2))\phi(\bm r_3)\right\rangle\!=\!C((\bm r_2\!-\!\bm r_1)^2, (\bm r_3\!-\!\bm r_1)^2, (\bm r_3\!-\!\bm r_2)^2),\nonumber
\end{eqnarray}
where $C(x, y, z)$ is a symmetric function of the three independent (dummy) variables, $x$, $y$ and $z$. We assumed spatial homogeneity and isotropy, which imply that the above correlations depend only on the geometry of the triangle whose vertices are $\bm r_i$. Consider next the tensor 
\begin{eqnarray}&&\!\!\!\!\!\!\!\!\!
D_{kl}(\bm r)\equiv \left\langle v_k(0)v_l(0)\phi(\bm r)\right\rangle
\nonumber\\&&\!\!\!\!\!\!\!\!\!
=\nabla_{1k} \nabla_{2l}
C(r_{21}^2, r_{31}^2, r_{32}^2)|_{\bm r_1=\bm r_2=0,\ \ \bm r_3=\bm r},
\end{eqnarray}
which is subsequently given by:
\begin{eqnarray}&&\!\!\!\!\!\!\!\!\!
D_{kl}(\bm r)=4 r_lr_{k} C_{yz}(0, r^2, r^2)-2\delta_{kl}C_x(0, r^2, r^2). \label{derivatives}
\end{eqnarray}
Taking the trace of the above equation,  introducing $q=r^2$, and using Eq.~(\ref{ts}) yields
\begin{eqnarray}&&\!\!\!\!\!\!\!\!\!
2q C_{yz}(0, q, q)-d C_x(0, q, q)\!=\!-\frac{\langle \psi(0)\phi(0)\rangle}{2}-\frac{\epsilon q}{4d},\end{eqnarray}
which is valid to linear order in $q$ since Eq.~(\ref{ts}) holds to quadratic order in $r$. Evaluating the above equation and its derivative with respect to $q$ at $q=0$, and employing the symmetry of $C(x, y, z)$, we obtain 
\begin{eqnarray}&&\!\!\!\!\!\!\!\!\!
C_x(0, 0, 0)\!=\!\frac{\langle \psi(0)\phi(0)\rangle}{2d},\ \ C_{xz}(0, 0, 0)\!=\!\frac{\epsilon}{8d(d\!-\!1)}.
\end{eqnarray}
In addition, use has been made of the fact that $C(x, y, z)$ must have finite cross derivatives of second order (in order to represent finite correlations of the flow) so that asymptotic expansions to the considered order can be obtained 
by differentiation. The above equation cannot be used at $d=1$ since in that case the three-point correlation function of $\phi(\bm r)$ depends on two, rather than three variables, thus leading to different considerations. 

We find that to linear order in $q$ (i.e. quadratic order in $r$) 
\begin{eqnarray}&&\!\!\!\!\!\!\!\!\!
C_x(0, q, q)=C_x(0, 0, 0)+q\left(C_{xy}(0, 0, 0)+C_{xz}(0, 0, 0)\right)
\nonumber\\&&\!\!\!\!\!\!\!\!\!
=\frac{\langle \psi(0)\phi(0)\rangle}{2d}+
\frac{\epsilon q}{4d(d-1)}. 
\end{eqnarray}
We therefore conclude by using the above in 
Eq.~(\ref{derivatives}) that to quadratic order in $r$ 
\begin{eqnarray}&&\!\!\!\!\!\!\!\!\!
D_{kl}(\bm r)=
\frac{\epsilon r_lr_{k}}{2d(d-1)} 
-\frac{\langle \psi(0)\phi(0)\rangle \delta_{kl}}{d}
-\frac{\epsilon r^2\delta_{kl}}{2d(d-1)}. 
\end{eqnarray}
In the next step towards our goal, we find that the third order correlation tensor of the velocity  $T_{ikl}(\bm r)$, which is symmetric in the last two indices, is given by:
\begin{eqnarray}&&\!\!\!\!\!\!
T_{ikl}(\bm r)\equiv \left\langle v_k(0)v_l(0)v_i(\bm r)\right\rangle=\nabla_iD_{kl}(\bm r)=-\frac{\epsilon r_i\delta_{kl}}{d(d-1)}\nonumber\\&&\!\!\!\!\!\!
+\frac{\epsilon\left( r_k\delta_{il}+r_l\delta_{ik}\right)}{2d(d-1)}
.\label{remarkable}
\end{eqnarray}
This is a remarkable result that is the counterpart of a similar result for incompressible turbulence. The only difference is in the numerical value of the coefficients, cf. \cite{frisch}. This difference however is qualitative and not only quantitative. The above form of $T_{ikl}(\bm r)$ gives  
\begin{eqnarray}&&\!\!\!\!\!\!\!\!\!\!\!\!\!\!
\left\langle \left(v_i(\bm r)-v_i(0)\right)\left(v_k(\bm r)-v_k(0)\right)\left(v_l(\bm r)-v_l(0)\right)\right\rangle
\nonumber\\&&\!\!\!\!\!\!\!\!\!\!\!\!\!\! %=-\langle v_i(\bm r)v_k(\bm r)v_l(0)\rangle
%-\langle v_i(\bm r)v_k(0)v_l(\bm r)\rangle+
%\langle v_i(\bm r)v_k(0)v_l(0)\rangle
%\nonumber\\&&\!\!\!\!\!\!\!\!\!\!\!\!\!\!
%-\langle v_i(0)v_k(\bm r)v_l(\bm r)\rangle+\langle v_i(0)v_k(\bm r)v_l(0)\rangle+\langle v_i(0)v_k(0)v_l(\bm r)\rangle\nonumber\\&&\!\!\!\!\!\!\!\!\!\!\!\!\!\!
=2\left(T_{lik}(\bm r)+T_{kli}(\bm r)+T_{ikl}(\bm r)\right)=0,
\end{eqnarray}
that holds at any $d$. This implies that all third order structure functions, both longitudinal and transversal ones, are identically zero. This is in sharp contrast with incompressible turbulence where the third order structure function, that describes the energy flux associated with the energy cascade \cite{frisch}, is non-zero both for $d=2$ as well as $d=3$. Here however, the third order structure function does not provide information on the cascade of the energy $\int \rho v^2 d\bm r/2$. It is rather a degenerate quantity that vanishes due to the effective one-dimensionality of the shocks. The direction of the shock formation is random, so that averaging over those directions the odd third power of velocity difference yields zero. 

Despite that the structure function vanishes, the third order correlation tensor of the velocity obeys the non-trivial Eq.~(\ref{remarkable}). That equation is predicted to hold in the supersonic inertial range and can be used for experimental testing of the reduction to Burgers turbulence.

The above considerations should be understood not as a statement on the vanishing of third order correlations of the velocity difference. They rather mean that the third order structure function of the velocity is of order higher than linear in $r/L$.
Furthermore, anisotropy, that is invariably present in various applications and is neglected above, could produce in the third order structure functions a contribution that would be larger than the isotropic value. The study of this possibility is left for future work. 

The above raises the issue of how to define the scaling exponents of the velocity, given that the third order structure function vanishes. The quantity that can be used for that purpose is the absolute value of the difference and not its longitudinal component, 
\begin{eqnarray}&&\!\!\!\!\!\!\!\!\!\!\!\!\!\!
\left\langle |\bm v(\bm r)-\bm v(0)|^p\right\rangle\propto r^{\zeta_p}. 
\end{eqnarray}
This is in contrast with the three-dimensional incompressible turbulence where scaling exponents defined with the help of longitudinal, transversal structure functions or the modulus of the difference would be identical, cf. \cite{Iyer}.  

{\bf Cascade direction at $Ma\gg 1$ in the potential case}---Reduction to Eqs.~(\ref{ns}) at $Ma\gg 1$ allows to understand the direction of the cascade in this limit. We first consider the case of potential Burgers' turbulence. Free flow evolution, obtained by discarding the acceleration term in Eqs.~(\ref{ns}), is described by the Burgers equation $\partial_t\bm v+(\bm v\cdot\nabla)\bm v=\nu \nabla^2\bm v$. In the case of potential flow this equation is integrable. It can be reduced to the linear heat equation by the Cole-Hopf substitution, see e.g.  \cite{Burgulence}. The solution demonstrates that in the small viscosity limit, the evolution consists of steepening of large scale profiles of the flow, which forms shocks with non-trivial geometry. This can be understood by observing that at small viscosity, Burgers' equation reduces to the Hopf equation $\partial_t\bm v+(\bm v\cdot\nabla)\bm v=0$, that governs the evolution of the velocity field of free particles. The evolution behaves according to the well-known implicit solution $\bm v(\bm x+\bm v_0(\bm x)t, t)=\bm v_0(\bm x)$ that express the conservation of the initial velocity $\bm v_0(\bm x)$ of a particle located initially at $\bm x$. Generally, linear trajectories of free particles would intersect at a finite time, making the velocity field multi-valued. The intersection is preceded by the blowup of $\nabla \bm v$ at the formation of an infinitely thin shock. Due to the singularity, the neglect of the viscous, higher derivative term in the Burgers equation becomes inconsistent near the blowup. Finite viscosity, however small it is, regularizes the divergences, arrests overturning of the shocks, and keeps the flow single-valued and smooth. The above implies that the direction of cascade in the inertial range, where by definition the forcing and viscosity terms are negligible \cite{frisch}, is direct because the flow generates smaller scale, infinitely thin, structures, that are eventually dissipated by the viscosity. Thus shocks of Burgers turbulence exhibit dissipative anomaly with energy dissipation remaining finite in the zero viscosity limit \cite{Burgulence}. This is in contrast to the lack of such anomaly in the case of the inverse cascade, see e.g. \cite{review}. 

In the presence of forcing then, the steady state consists of continuous injection by the force of flow profiles of scale of order $L$, which then generate smaller scale structures until dissipated by viscosity at the smallest scale of the flow. The cascade of the Burgers turbulence is direct in any dimension, in contrast to the incompressible Navier-Stokes equations. 

It must be remarked that here we speak of the cascade in the sense of generation of higher wave numbers of the Fourier image of velocity 
field, since $\int v^2 d\bm x/2$ is not conserved and cannot be regarded as energy. In the absence of external force and dissipative processes, Eqs.~(\ref{ns}) imply $\partial_t \rho v^2+\nabla\cdot \left(\rho v^2\bm v\right)=0$ so that the energy which is conserved is the kinetic energy $\int \rho v^2 d\bm x/2$. Since the density $\rho$ is passive it is a functional of the velocity, so that $\rho v^2$ must be considered as energy defined in terms of the velocity field only. 

{\bf Generalization to non-potential stirring}---If the accelerations' field is non-potential then the flow has finite vorticity, generated by the source term, $s$, in Eq.~(\ref{fp}). In this case the flow at $Ma\to\infty$ still obeys the Burgers equation, given by the last of Eqs.~(\ref{ns}). However, the equation with $\bm a=0$ is no longer integrable to the best of our knowledge. The same Hopf equation of free particles arises at zero viscosity, which implies that the cascade is still direct. This conclusion is the extension of observations of \cite{fk} that were described in the introduction. Thus, as the Mach number increases from zero the kinetic energy transfer to smaller scales from $L$ increases monotonously. Generally this process may be described by the order parameter $R$ that measures the ratio of the kinetic energy fluxes from the forcing scale to small and large scales.
  The results of \cite{fk} indicate that $R(Ma)$, considered as a function of the Mach number, increases monotonously and continuously from $R(0)=0$ to $R(\infty)=\infty$ with $R(Ma\sim 1)\sim 1$. Thus the transition from inverse to direct cascade is not sharp, it is continuous. 
  %AK: Correct.
  The degree of universality of $R(Ma)$, i.e., its dependence on the forcing is of interest.

%
%\begin{figure}[ht]
% \centerline{\includegraphics[scale=0.32]{den-snap2_00154}}% Images in %100% size
% \caption{Density}\label{Fig:M01}
%\end{figure}
%
%\begin{figure}[ht]
% \centerline{\includegraphics[scale=0.32]{den-snap2_04000}}% Images in %100% size
% \caption{Density}\label{Fig:M06}
%\end{figure}
%
%\begin{figure}[ht]
% \centerline{\includegraphics[scale=0.32]{den_0100}}% Images in 100% %size
% \caption{Density}\label{Fig:M096}
%\end{figure}

{\bf Lower cutoff scale of the cascade}---We recall that the direct cascade of incompressible three-dimensional turbulence ends at the lower cutoff scale determined by the viscosity \cite{frisch}. In our case, the direct Burgers-type cascade of velocity variance (specific kinetic energy) would typically end at the sonic scale $l_s$. Comparing the material derivative and pressure terms in the last of Eqs.~(\ref{nsp}), we see that the definition of $l_s$ implies that at this scale the neglected "small" pressure term in the last of Eqs.~(\ref{nsp}) is of the same order as the material derivative term. Hence at scales smaller or of order $l_s$ the pressure forces must be included. In contrast, typically viscosity is irrelevant at $l_s$, becoming significant only at a much smaller scale. Hence what changes the direct Burgers-type cascade downscales from $L$ is the pressure at $l_s$.  

The structure of the flow below $l_s$ can be understood by considering it as an effective flow with Mach number of order one. (The local, scale-dependent Mach number is defined as the typical velocity of eddies at that scale divided by $c_s$, see \cite{sonic} and cf. with local Reynolds number \cite{frisch}. By definition, the local Mach number at $l_s$
is of order one.) Here all scales larger than $l_s$ provide effective random forcing for the smaller scales' flow (this is only a qualitative statement since the small scales react back on the large scales). Below $l_s$ the eddies have subsonic velocities, and for some aspects the turbulence is effectively incompressible. For instance, as already said, the density $\rho$ is roughly constant below $l_s$, as in the incompressible case \cite{sonic,dust,fm}. 

{\bf Kinetic energy transfer between Fourier modes at $Ma\lesssim 1$}---We study the nature of turbulence at $Ma\lesssim 1$ by considering how incompressible turbulence, that holds at $Ma=0$, is changed by a small but finite $Ma$. For finite $Ma$ the enstrophy is no longer an integral of motion of two-dimensional Navier-Stokes equations without forcing and dissipation. Turbulence then can be thought of as arising from complex interactions of planar vortices and sound waves, where vortices tend to support 
%AK: maybe "tend to support ..."
inverse and waves direct energy cascades (this way of thinking might need refinements when temperature degree of freedom is present) \cite{fk}. At small but finite $Ma$ the vortices still produce inverse cascade of energy and the energy injected into the flow at the pumping scale $L$ flows to larger scales. That inverse cascade is associated with velocity magnitude growth as the cascade reaches larger and larger scales. Characteristic velocity $v_l$ of eddies with scale $l$ obeys the Kolmogorov scaling $v_l\sim u (l/L)^{1/3}$, that is valid for two-dimensional inverse cascade \cite{review}. That scaling however needs to be modified to $v_l\sim u (l/L)^{1/2}$, that corresponds to $k^{-2}$ spectrum, for $l$ so large that $v_l$ is comparable with the speed of sound \cite{fk}. When the cascade reaches the sonic scale $l_s\sim Ma^{-3} L$ (this estimate is done using Kolmogorov scaling, the actual law depends on lengths of different scaling ranges and can be readily worked out), by definition, the characteristic flow velocity equals speed of sound. Here we assume that the system size is larger than $l_s$ and observe that $l_s\gg L$ holds at even not so small $Ma$. At $l_s$ ``resonance'' condition is obeyed and turbulence starts to emit sound waves strongly (of course the emission happens also at intermediate scales $L\lesssim l\lesssim l_s$, however in that range it is too weak to stop the cascade). The waves then transfer energy to small scales, thereby creating an energy flux loop observed in \cite{fk}. Thus, the energy is predominantly contained at scales $l_s$ whereas its dissipation occurs at smallest scales generated by the shock waves of the direct cascade.

The density associated with the above cascade is a large-scale field that has significant variations only at scales of order $l_s$, see above and Fig.~\ref{Fig:M06}. The smallest scales are also present in the density however their contribution to the spatial density profile is negligible (this is not true for high-order derivatives of the density however). 

Physics at $Ma\sim 1$ can then be understood by asymptotic continuation of the $Ma\ll 1$ case. When we consider larger, yet small $Ma$, the sonic scale $l_s$ decreases and becomes of order $L$ at $Ma\sim 1$. This is the situation where the characteristic velocity injected by the driving force equals the speed of sound. The largest scale of the energy flux loop, introduced at $Ma\ll 1$, is then of order of the pumping scale $L$. Vortices still produce some kinetic energy flux up the hierarchy scals,
%AK: avoid jargon -- "up the hierarchy of scales"
however it does not reach scales much larger than $L$, see \cite{fk}. The waves take energy to smaller scales from scales of order $L$, so that two-dimensional turbulence at $Ma\sim 1$ exhibits essentially only the direct energy cascade, if not to account for significant spectrum at wavenumbers somewhat larger than $L^{-1}$. Small-scale turbulence in this case is still subsonic and contains vortices which decelerate the direct cascade however do not stop it. Since subsonic eddies of the small-scale turbulence are effectively incompressible, they do not create significant spatial variations of $\rho$. Density is still a large-scale field with variations at scale $l_s\sim L$, similarly to $Ma\ll 1$ case, see above and Fig.~\ref{Fig:M01}.

{\bf Density multifractality in supersonic inertial range}---We turn now to discuss cases with $Ma \gg 1$. The fact that in that regime the density field is a passive scalar that is governed by 
Eqs.~(\ref{ns}) has far reaching implications. It was demonstrated that, provided that the scalings of solenoidal and dilatational components of the flow are similar, the density field generated by the continuity equation is multifractal \cite{fm}. 
%AK: See https://arxiv.org/abs/1802.08228 , where I discussed scaling in 2D simulations; this work can potentially be useful for your subsection on scaling.
At Mach number values that are much smaller than $1$ the scaling exponents of those velocity components are not equal, however, as the Mach number grows, the difference between them decreases \cite{fk,K2018}. When it reaches small enough values a multifractal density structure holds \cite{fm}.

We first consider the potential case. The scaling condition above pertains to scaling exponents, and not to the absolute magnitude of the velocity components, so in the potential case it holds trivially. Then, the results of \cite{fm} imply that in the potential case the density field is multifractal in the supersonic inertial range while below the scale $l_s$ there are no appreciable fluctuations of the density. We here confine ourselves with implications for the density spectrum $E(k)$, that obeys for $L^{-1}<k<l_s^{-1}$ the power-law $E(k)\propto k^{\delta-1}$ with $\delta>0$. The so-called correlation dimension of the multifractal is defined as $2-\delta$, and is strictly smaller than the full space dimension two, see \cite{kritsuk} for observations in the three-dimensional case. The density fluctuations obey in that case
\begin{eqnarray}&&\!\!\!\!\!\!\!\!\!\!\!\!\!\!
\left\langle \rho^2\right\rangle=\int_0^{\infty} E(k)dk\simeq C_0\left(\frac{L}{l_s}\right)^{\delta}\simeq C_0C^{\delta}Ma^{2\delta}, \label{fls}
\end{eqnarray}
where $C_0$, $C$ are constants and we assumed that the leading order behavior of the integral is determined by the power-law cutoff at $k\sim l_s^{-1}$. We assumed the Burgers' $k^{-2}$ spectrum in estimating $L/l_s$ as $C Ma^2$ with a constant $C$, cf. above. The fluctuations diverge at $Ma\to\infty$ where the cutoff scale of the multifractal $l_s$, at which the density is supported, tends to zero.

%\begin{comment}
%This implies that moments of the density obey
%\begin{eqnarray}&&\!\!\!\!\!\!\!\!\!\!\!\!\!\!
%\left\langle \rho^p\right\rangle=C_p\left(\frac{L}{l_s}\right)^{\xi(p)}=C_p C^{2\xi(p)} Ma^{2\xi(p)},
%\end{eqnarray}
%where the positive exponents $\xi(p)$ are non-trivially related to the spectrum of fractal dimensions \cite{fm}. Here $C_p$ are numerical constants and 
%\end{comment}
%We observe that due to rotational and parity symmetries
%\begin{eqnarray}&&\!\!\!\!\!\!\!\!\!\!\!\!\!\!
%\left\langle v_i(0)v_k(\bm r)v_l(\bm r)\rangle=A(r){\hat r}_i
%{\hat r}_k{\hat r}_l+B(r)\left(\delta_{ik}{\hat r}_l+\delta_{il}{\hat r}_k+\delta_{kl}{\hat r}_i\right),
%\end{eqnarray}
%which implies that 
%\begin{eqnarray}&&\!\!\!\!\!\!\!\!\!\!\!\!\!\!
%\langle v_i(0)v^2(\bm r)\rangle=A(r){\hat r}_i
%+B(r)(d+2){\hat r}_i,\ \ 
%\left\langle v_i(0)v_k(\bm r)v_l(\bm r)\rangle
%\end{eqnarray}
%\begin{eqnarray}&&\!\!\!\!\!\!\!\!\!\!\!\!\!\!
%\langle \left(v_i(\bm r)-v_i(0)\right)
%\left(\bm v(\bm r)-\bm v(0)\right)^2\rangle=
%-2\langle v_i(\bm r)v_k(\bm r)v_k(0)\rangle
%\nonumber\\&&\!\!\!\!\!\!\!\!\!\!\!\!\!\!
%-2\langle v_i(0)v^2(\bm r)\rangle+2\langle v_i(0)v_k(\bm r)v_k(0)\rangle=
%\end{eqnarray}

In the case of non-potential acceleration field, scalings of the solenoidal and dilatational 
components in the supersonic inertial range are identical, for generic acceleration fields. This is because the evolution in that range is describable by the Hopf equation $\partial_t\bm v+(\bm v\cdot\nabla)\bm v=0$ that does not induce difference in the components' scalings, cf. above. Then we find again Eq.~(\ref{fls}). 

Thus supersonic two-dimensional turbulence is not that different from its three-dimensional counterpart, in sharp contrast to the incompressible limit of small $Ma$, cf. \cite{dust,fm}. Both exhibit transfer of energy from large to small scales and multifractal density in the supersonic inertial range. 

{\bf Burgers' model for accretion}---The above considerations suggest the introduction of a minimal model for accretion that is expected to capture the main physics of supersonic thin disks. The model relies on the fact that the velocity is a large-scale field and the density is smooth below $l_s$. We use as a model the same Eqs.~(\ref{ns}) however with the value of viscosity such that the viscous scale is $l_s$. Thus the proposed velocity equation is
\begin{eqnarray}&&\!\!\!\!\!\!\!\!\!\!\!\!\!\!
\partial_t\bm v+(\bm v\cdot\nabla)\bm v=\bm a+\frac{b}{Ma^3}\nabla^2\bm v,\label{proposed}
\end{eqnarray}
where $b$ is the adjustment numerical factor of order one which is considered below. The coefficient of the Laplacian term is such that the ratio of the non-linear and viscosity terms is of order one at $l_s$. Here for estimates we assume velocity scaling that corresponds to $k^{-2}$ scaling of the spectrum. The density obeys a modified continuity equation (diffusion equation in moving fluid) 
\begin{eqnarray}&&\!\!\!\!\!\!\!\!\!\!\!\!\!\!
\partial_t\rho+\nabla\cdot(\rho\bm v)=\frac{b}{Ma^3}\nabla^2\rho.\label{proposa}
\end{eqnarray}
Here we observe that since the model omits the pressure term altogether, then continuity equation without diffusivity could produce appreciable density variations below $l_s$, as in the so called Batchelor regime \cite{review}. Introduction of the Laplacian term into Eq.~(\ref{proposa}) is a means to obtain the required density smoothness below $l_s$. 

It is expected that Eqs.~(\ref{proposed})-(\ref{proposa}) reproduce rather accurately the flow in the supersonic inertial range, which is the main range of interest in the accretion problem (this is the range where appreciable density variations occur and also the range that determines the overall velocity and kinetic energy). The advantage of this model is that in the case of potential accelerations, which is relevant for gravity-driven accretion, the velocity equation is mapped into a linear equation by the Cole-Hopf transformation $\bm v=-2b Ma^{-3}\nabla \ln f$. We have ($\bm a=-\nabla\psi$)  
\begin{eqnarray}&&\!\!\!\!\!\!\!\!\!\!\!\!\!\!
\partial_tf=\frac{Ma^3}{2b}\psi     f+\frac{b}{Ma^3}\nabla^2f.
%\nonumber\\&&
%\partial_t\bm v=-\nabla \psi +2b Ma^{-3}\nabla^2\nabla \bm v
\end{eqnarray}
It is much easier to perform simulations of the above (imaginary time Schrodinger) linear equation than of the original non-linear equations. We conjecture that the above model would faithfully reproduce the accretion rate by an appropriate choice of $b$. On the theoretical side, the linear model suggests that supersonic potential turbulence is much more stable than turbulence usually is. 

The model miscalculates possibly large fluctuations of the supersonic cascade's cutoff scale $l_s$ that exist due to intermittency. Thus the original cutoff due to pressure is different from the model's cutoff due to viscosity. For $Ma\lesssim 10$ intermittency is moderate and this seems to be not of much concern. However, for the highest Mach number of practical relevance, $Ma\sim 100$ we have $L/l_s\sim 10^4$. The same length of inertial interval in three-dimensional incompressible turbulence corresponds to Reynolds numbers of order $10^5-10^6$ which are appreciably intermittent \cite{frisch}. The degree to which the proposed model is accurate must be studied numerically and is left for future work. 
 
{\bf Magnetohydrodynamics}---
%Any solution of the hydrodynamic equations must be checked for stability. For conducting fluids, this makes the above description of the pure hydrodynamic steady state incomplete. For these fluids,invariably present 
Small fluctuations of magnetic field, invariably present in astrophysical environments, may be significantly  amplified by the turbulence.
%, demanding the field's inclusion into the system of hydrodynamic equations. 
In such cases the hydrodynamic equations need to be extended in order to include the influence of the magnetic fields. This is the case for three-dimensional turbulence, see e.g. \cite{ll8}, which is believed to be the reason for the ubiquitous magnetic fields in astrophysics. Here we demonstrate that in the two-dimensional case the situation is quite similar. However the source of the magnetic field which is amplified by turbulence is not small fluctuations, which in fact decay, but rather large-scale magnetic field. 

It is often believed that the magnetic field is irrelevant in the two-dimensional situation due to the antidynamo theorem \cite{antidynamo}, which we consider here in more detail. The theorem states that there is no dynamo effect in incompressible two-dimensional turbulent flow as any magnetic field inevitably decays to zero. Thus, it was observed that the vertical component of the magnetic field obeys the heat conduction equation in a moving fluid \cite{antidynamo}. This implies that this component of the field relaxes to a constant, similarly to how density of tracers becomes uniform due to mixing by incompressible turbulence. Correspondingly, the vertical component disappears from the condition of magnetic field's solenoidality. The horizontal components can then be described by a scalar potential, that again obeys the heat conduction equation. Therefore, the potential decays to a constant, which implies vanishing of the horizontal components.
Thus, for incompressible flow the steady state solution is a constant vertical field that generates no Lorentz force and exerts no influence on the flow \cite{antidynamo}. 

Despite some doubts \cite{gk}, the antidynamo theorem for incompressible flow has been demonstrated to be valid (see for instance \cite{igor}) and appears today as a solid result whose universality has far-reaching implications. Thus, in the case of two-dimensional incompressible flows, the consideration that assumes that there are no magnetic fields in the turbulent flow of conducting fluid is self-consistent: small fluctuations would asymptotically decay with time. These fluctuations can be three-dimensional: the Lorentz force is quadratic in the magnetic field. Thus small three-dimensional fluctuations exert negligible effect on the flow which remains two-dimensional with good approximation. Still care is needed since significant transient growth is possible, see e.g. \cite{igor,tr}, which could temporarily turn on the interaction of the magnetic field with the flow. 

In fact, the antidynamo theorem shows that any fluctuations of the magnetic field, be they large or small, whether they interact with the flow for some time or not, decay asymptotically on one condition: the Lorentz force due to the magnetic field must keep the incompressible flow (approximately) two-dimensional. Hence there remains the possibility for non-linear instability of incompressible flow, where a large fluctuation of magnetic field would create the third component of velocity that will prevent the fluctuation's decay. Summarizing, the antidynamo theorem for incompressible flows leads to the conclusion that, if the flow can be considered two-dimensional, then the steady state magnetic field is zero. 

Any real flow, however, has finite compressibility, so the above anti-dynamo theorem, at best, describes the case of $Ma\ll 1$, where the flow is approximately incompressible. There is currently no result similar to Zeldovich theorem that allows us to assess the relevance of compressibility to sustaining non-trivial steady-state magnetic fields. Moreover, there is no demonstration that at $Ma\ll 1$ the neglect of the magnetic field is indeed valid (we will show that this is not always the case). In this section we fill that gap and explore the role of flow compressibility in the magnetic field amplification process. 

We study the transport of a solenoidal magnetic field $\bm B$ by a two-dimensional flow $\bm v$. We use the same framework as in the Zeldovich antidynamo theorem \cite{antidynamo}, relaxing the condition of incompressibility. Hence the study below can be considered as compressible generalization of the antidynamo theorem in two dimensions. The evolution of magnetic field $\bm B$ is described by the magnetohydrodynamic equations \cite{ll8},
%\begin{eqnarray}&&\!\!\!\!\!\!\!\!\!\!\!\!\!\!
%\partial_t\bm B=\nabla\times(\bm v\times\bm B) +\eta\nabla^2 \bm B,\ \ \nabla\cdot \bm B=0.
%\end{eqnarray}
%The first equation can be written as,
\begin{eqnarray}&&\!\!\!\!\!\!\!\!\!\!\!\!\!\!
\partial_t\bm B+(\bm v\cdot \nabla) \bm B=(\bm B\cdot \nabla) \bm v-\bm B\nabla\cdot \bm v+\eta\nabla^2 \bm B, \label{magn}
\end{eqnarray}
while the flow velocity $\bm v$ obeys 
\begin{eqnarray}&&
\partial_t\bm v+(\bm v\cdot\nabla)\bm v=-\frac{1}{Ma^2}\nabla \ln \rho+\bm a+\nu\nabla^2\bm v\nonumber\\&&
-\frac{1}{2\rho}\nabla B^2+\frac{(\bm B\cdot\nabla)\bm B}{\rho}, \label{lorne}
%+\frac{\nabla_k\sigma'_{ik}}{\rho}, 
\end{eqnarray}
where the last two terms describe the Lorentz force, cf. Eqs.~(\ref{nsp}). The density obeys the continuity equation given by the first of Eqs.~(\ref{nsp}). The velocity field is considered to depend only on the horizontal coordinates $x$ and $y$ and to have zero vertical component, $v_z=0$. In contrast, we consider a fully three-dimensional magnetic field that generally has a vertical component $B_z$, and that depends on $x$, $y$ and $z$. We impose constraint on the magnetic field that the vertical component of the flow and the $z-$dependence of $\bm v$, as generated by the Lorentz force, are negligible. They must be negligible to the extent that the magnetic field lines' stretching term $(\bm B\cdot \nabla) \bm v$ in the equation for $B_z$, which vanishes for planar $\bm v$ identically, is negligible. This produces the following closed equation for the vertical component of $\bm B$
\begin{eqnarray}&&\!\!\!\!\!\!\!\!\!\!\!\!\!\!
\partial_t B_z\!+\!\nabla \!\cdot  \!( B_z\bm v) \!=\!\eta\nabla^2  B_z.\label{sgo}
\end{eqnarray}
We shortly recapture the Zeldovich antidynamo theorem that holds in the case of incompressible flow. 

{\bf There is no magnetic field in steady state for incompressible flow}---Eq.~(\ref{sgo}) implies for incompressible flow \begin{eqnarray}&&\!\!\!\!\!\!\!\!\!\!\!\!\!\!
\partial_t\int B_z^2 d\bm x=-2\eta\int \left(\nabla B_z\right)^2d\bm x,\label{so}
\end{eqnarray}
    where it it assumed that the boundary terms that are generated by integrating by parts can be neglected as in \cite{antidynamo}. Eq.~(\ref{so}) implies that magnetic energy decays until the field vanishes (the possible non-zero constant field can be eliminated by considering fluctuations around it). This conclusion on the vanishing of a steady-state $B_z$ holds irrespective of whether $B_z$ reacts on $\bm v$ via the momentum equation, it rests solely upon the assumption that $\bm v$ is two-dimensional \cite{antidynamo}.  

The vanishing of the horizontal components of the magnetic field for incompressible flow is proved similarly. Due to $B_z=0$, the condition that $\bm B$ is solenoidal yields $\partial_x B_x+\partial_y B_y=0$. We can therefore describe these components with the help of scalar potential $A$ according to $B_x=\partial_y A$ and $B_y=-\partial_x A$. That potential obeys an advection-diffusion equation
\begin{eqnarray}&&\!\!\!\!\!\!\!\!\!\!\!\!\!\!
\partial_t A\!+\!(\bm v\!\cdot\!\nabla) A\!=\!\eta\nabla^2 A, \label{adv}
\end{eqnarray}
which can be verified by taking derivatives and using $\partial_x v_x+\partial_y v_y=0$. We see that for incompressible flow $B_z$ and $A$ obey the same equation. By the same argument, $A$ relaxes to a constant, implying $B_x=B_y=0$. Thus the magnetic field for steady state incompressible planar flow is zero. 

{\bf Any planar flow of conducting fluid is magnetised in the limit of strong external field}---Incompressible flow has a degeneracy - submitting the system to an external magnetic field does not change the flow at all. The solution is given by the superposition of that external field and the solution that holds at zero field. Compressibility removes this degeneracy, inducing the flow to change in response to the imposed external field. (We will see below, that the impact of that field is determined by its total flux through the system.) As discussed above, Eq.~(\ref{sgo}) is solved by a constant in the incompressible case only. When the flow is compressible, writing $B_z$ as the sum of the constant external field $B^e$ and the internal field $B^i$ gives 
\begin{eqnarray}&&\!\!\!\!\!\!\!\!\!\!\!\!\!\!
\partial_t B^i\!+\!(\bm v\cdot\nabla) B^i \!=\!-B^e\nabla\cdot\bm v+\eta\nabla^2  B^i.\label{generat}
\end{eqnarray}
The constant external field generates an inhomogeneous source in the equation for the internal field. In turn, that creates a non-trivial Lorentz force in Eq.~(\ref{lorne}) that  changes the velocity field. Thus imposing on the system an external magnetic field can be used to manipulate the flow externally. 

The above demonstrates that at $Ma\ll 1$ using a large external field, finite or large effect on the flow can be created via $B^i$. In this case, anticipating the vanishing of the horizontal components of the magnetic field proved below, Eq.~(\ref{lorne}) becomes 
\begin{eqnarray}&&\!\!\!\!\!
\partial_t\bm v\!+\!(\bm v\!\cdot\!\nabla)\bm v\!=\!-\frac{\nabla \ln \rho}{Ma^2}\!+\!\bm a\!+\!\nu\nabla^2\bm v
\!-\!\frac{B^e}{\rho}\nabla B^i,\label{ms}
%+\frac{\nabla_k\sigma'_{ik}}{\rho}, 
\end{eqnarray}
where we assume $B^e\gg B^i$. We consider as an example, the pseudosound mode dominated regime of $Ma\ll 1$ where velocity divergence scales as $Ma^2$, see e.g. \cite{pseudos,levermore}. We see then that taking $B^e\propto Ma^{-1}$ we find $B^i\propto Ma$ from Eq.~(\ref{generat}). This leads to magnetic pressure force in Eq.~(\ref{ms}) that is of order one and has a finite effect on the flow (note that the assumption $B^e\gg B^i$ is correct). Furthermore, $B^e\gg Ma^{-1}$ would produce a large Lorentz force, leading to complete rearrangement of the flow.  

A remark is in order. We assumed that at $Ma\ll 1$ the flow divergence is $Ma^2$ times a field of order one. In fact, this is true in the three-dimensional case \cite{pseudos,levermore}, however, might need refinements in two dimensions. This is because of the inverse cascade that could cause $\nabla\cdot\bm v$ to be a field whose dominant modes are of order of $l_s^{-1}$. From the low-Mach number spectrum of the dilatational component of the velocity, $E_d(k)$, presented in \cite{fk} this seems to be not the case. 
%AK: https://arxiv.org/abs/1802.08228 discusses the spectrum of div(u) and therefore should be relevant here too.
It is seen by using that the spectrum of $\nabla\cdot\bm v$ is $k^2 E_d(k)$ that divergence is determined by eddies at the pumping scale. Thus it appears that the above considerations hold also in the two-dimensional case. However, the spectrum of $\nabla\cdot\bm v$ presented in \cite{K2018} seems to indicate that the divergence is a small-scale field. Further studies are necessary, however, even if another factor should be present in $\nabla\cdot\bm v$, this does not change our main conclusion - any flow of a conducting fluid in large enough magnetic field will generate appreciable Lorentz force and hence evolve differently from the pure hydrodynamic case. In other words, in the presence of non-zero magnetic flux compressibility is a singular perturbation 
%AK: maybe rather incompressible turbulence is a singular limit when magnetization is present?
-- even when it is small it can have a finite or large effect on the flow when compensated by a large external magnetic field. We leave detailed investigation of the critical value of $B^e$, above which the flow changes at $Ma\ll 1$, for future work.

{\bf Stationary measure $n_s(t, \bm x)$}%
%AK: is it essential to have n(t,x) instead of more traditional n(x,t) here and below?
---We saw that inclusion of compressibility might introduce a qualitative change in the flow by allowing for the magnetic degrees of freedom to affect the steady state. Eq.~(\ref{so}) becomes 
\begin{eqnarray}&&\!\!\!\!\!\!\!\!\!\!\!\!\!\!
\partial_t\int B_z^2 d\bm x=-\int  B^2\nabla\cdot\bm v
-2\eta\int \left(\nabla B_z\right)^2d\bm x,
\end{eqnarray}
and we can no longer apply the argument of \cite{antidynamo}, i.e. compressible planar flow may sustain a finite magnetic field in the steady state. Insight into the structure of the stationary magnetic field is gained by comparing it to a passive concentration field $n(t, \bm x)$ that obeys
\begin{eqnarray}&&\!\!\!\!\!\!\!\!\!\!\!\!\!\!
\partial_t n\!+\!\nabla \!\cdot  \!( n\bm v) \!=\!\eta\nabla^2  n,\ \ 
\int nd\bm x=1. \label{sns}
\end{eqnarray}
The difference from Eq.~(\ref{sgo}) is that $n$ is positive, does not react on $\bm v$ and has finite unit mass. Despite these differences, it is anticipated that $B_z$ in the steady state is proportional to the steady state solution $n_s$ of the above equation, as will be proved rigorously below. 

Eq.~(\ref{sns}) coincides with the equation governing the distribution of particles that diffuse within a moving fluid. At zero diffusion, $\eta=0$, it has the same form as the continuity equation whose density solution is steady. We will assume below that Eq.~(\ref{sns}) in cases of interest generates a steady state solution. This solution, the so-called stationary measure $n_s(t, \bm x)$, can be derived as the infinite evolution time limit 
\begin{eqnarray}&&\!\!\!\!\!\!\!\!\!\!\!\!\!\!
n_s(t, \bm x)\equiv \lim_{t_0\to -\infty}n(t, \bm x),\ \ 
n(t_0)=n_0(\bm x). \label{sno}
\end{eqnarray} 
Above $t_0$ is the time at which the initial condition for Eq.~(\ref{sns}) is set, and it is assumed that $\int n_0(\bm x) d\bm x=1$. It will be seen below that the above equation defines a unique $n_s(t, \bm x)$ that is independent of $n_0(\bm x)$.

Steady state for Eq.~(\ref{sns}) can be realized by assuming that the flow is confined to a finite volume, as when the velocity field vanishes or is periodic on finite boundaries. Despite that these boundary conditions (b.c.) are not necessarily true, they provide a useful framework within which Eq.~(\ref{sns}) generates a steady state (roughly the particles cannot escape and their concentration reaches a steady state). These (artificial) b.c. will be explained in more detail below. 

Structure of $n_s(t, \bm x)$ can be understood by considering first the case of $\eta\to 0$ where $n_s(t, \bm x)=\rho(t, \bm x)$ could be anticipated. In fact, $n_s(\eta=0)=\rho$ is typically untrue, since the continuity equation is not dissipative and therefore the passive field $n$ and the active field $\rho$ may differ, see detailed discussion in \cite{fm}. Still, it is true that $n_s(\eta=0)$ is qualitatively similar to $\rho$. 

Finite $\eta$ smoothens $n_s(\eta=0)$ over the so-called resistive scale $l_{\rm res}$. This is the scale so small that at $l_{\rm res}$ the resistive term in Eqs.~(\ref{sgo}) or (\ref{sns}) is comparable with the advective term. Typical velocity $u_{\rm res}$ of eddies at scale $l_{\rm res}$ is such that the local magnetic Reynolds number $u_{\rm res}l_{\rm res}/\eta$ is of order one. The actual value of $l_{\rm res}$ is determined by using velocity scaling at larger scales, cf. the definition of the viscous scale \cite{frisch}, and is not necessary here. Thus roughly $n_s(\eta)$ is $n_s(\eta=0)$ smeared over the scale $l_{\rm res}$. Combining this with the previous observation of the similarity of $n_s(\eta=0)$ and $\rho$, we conclude that $n_s(t, \bm x)$ is qualitatively similar to $\rho(t, \bm x)$ coarse-grained over the scale $l_{\rm res}$ (this only means that the statistics are similar, and does not necessarily hold per realization). 

We conclude that at $l_{\rm res}\lesssim l_s$, where $l_{\rm res}$ is not larger than the smoothness scale of the density, the fields $n_s(t, \bm x)$ and $\rho(t, \bm x)$ have statistically similar properties. In contrast, if $l_{res}\gg l_s$ then the fields are quite different. In this case $n_s(t, \bm x)$ is a large scale field, smooth over $l_{\rm res}$, at $Ma\lesssim 1$. If $Ma\gg 1$ and $l_s\ll l_{\rm res}\ll L$ then $n_s(t, \bm x)$ is smooth below $l_{\rm res}$ and multifractal above it, until the scale $L$. Finally, at $Ma\gg 1$ and $l_{\rm res}\gtrsim L$ the field $n_s(t, \bm x)$ is a large scale field, smooth over the scale $l_{\rm res}$.

{\bf Magnetic field relaxes to $n_s(t, \bm x)$}---To study the magnetic field we need to introduce boundary conditions for Eq.~(\ref{sgo}). To understand the role of these conditions, we observe that Eq.~(\ref{sns}) also arises in the study of the motion of inertial particles in turbulence, already mentioned above (there usually the diffusion is neglected). The main motivation for the study of the particles' problem is prediction of the rain formation, see e.g. \cite{shaw,us}. We shortly review the relevant physics.

In warm clouds water droplets are located in a volume that does not have sharp boundaries. Yet, to study their distribution in the bulk of the cloud, they are usually represented as particles that move in a finite box. Due to the confinement to a finite volume, such particles necessarily reach a steady state distribution. Properties of this distribution are then applied to droplets in clouds, despite that they do not obey the same boundary conditions. The rationale for this approach is that relaxation to the steady state is a local process. Thus, droplets in a cloud redistribute locally according to the steady state in a box, and the actual effects of the boundary conditions take much longer times. Hence the boundary conditions of a finite box are used as merely a device to study the transient period when boundary phenomena do not penetrate the bulk. Here we use the same approach, using the boundary conditions that result in the steady state for Eq.~(\ref{sgo}) and then applying the result to the actual magnetic field locally, far from the disk boundaries. 

We wish to show that in steady state $B_z(t, \bm x)$ coincides with $n_s(t, \bm x)$. This is despite that, as already mentioned, generally, $B_z$ is active field that reacts on the velocity, and $n_s$ is a passive field that does not. The issue of whether an active and passive fields, obeying the same first order in time equation, coincide is quite old in turbulence. The most famous example is vorticity and passive magnetic field, see \cite{fm} for a recent detailed discussion and also \cite{review}. In our case, however, the number of degrees of freedom is so constrained that the active and passive fields obeying the same Eq.~(\ref{sgo}) do coincide in the long-time limit. Our demonstration of this below is an adaptation of \cite{fm}. 

Long-time properties of $B_z(t, \bm x)$ can be obtained by studying the Green's function $G(t, \bm x| \bm x')$ that obeys,
\begin{eqnarray}&&\!\!\!\!\!\!\!\!\!\!\!\!\!\!
\partial_{t}G(t, \bm x| \bm x')\!+\! \nabla \!\cdot  \! \left( G(t, \bm x| \bm x')\bm v(t, \bm x)\right) \!=\!\eta\nabla^2 G(t, \bm x| \bm x'),\label{green}
\end{eqnarray}
where $G(t=0, \bm x| \bm x')=\delta(\bm x-\bm x')$. We have $B_z(t, \bm x)=\int G(t, \bm x| \bm x') B_z(0, \bm x')d\bm x'$ where it is assumed that the Green's function obeys proper boundary conditions that guarantee that
$B_z(t, \bm x)$ also obeys them. We assume that finite box boundaries whose vertical dimension is much smaller than the horizontal dimensions can be used. The assumed details of the boundary conditions are simple to see from how they appear in the calculations. We set the volume of the domain of the flow to one by choice of units of length (here our dimensionless variables differ from those used in the study of velocity). We assume that, similarly to \cite{antidynamo}, natural boundary conditions for $G(t, \bm x| \bm x')$ can be used \cite{fp} so that the spatial integral of $G(t, \bm x| \bm x')$, and thus also of $B_z(t, \bm x)$, is conserved. Then $G(t, \bm x| \bm x')$ can be interpreted as the probability density function (PDF) of the coordinate $\bm x(t)$ of the Brownian particle given that $\bm x(0)=\bm x'$ (transition probability). The particle's motion is driven by combination of transport by the flow and diffusion with diffusion coefficient $\eta$. The main observation is that $G(t, \bm x| \bm x')$ becomes independent of $\bm x'$ at large times. For time-independent flow $\bm v$ this is the consequence of the PDF's relaxation to the time-independent stationary solution. For time-dependent random stationary flows, of interest here, $G(t, \bm x| \bm x')$ becomes at large times a stationary random distribution $n_s(t, \bm x)$ that describes the concentration of ensemble of large number of independently moving Brownian particles (or the PDF of one particle), introduced previously.

We demonstrate that at large times $G(t, \bm x| \bm x_1)=G(t, \bm x| \bm x_2)$ holds for any $\bm x_1$ and $\bm x_2$. We use the entropy functional approach \cite{fp} which here serves as the counterpart of studying the evolution of magnetic energy in \cite{antidynamo}, see Eq.~(\ref{so}). We observe that $G(t, \bm x| \bm x')$ is strictly positive at any $t>0$ as can be seen, e.g., from the path integral representation of the transition probability, provided below. Thus at $t>0$ we can define the distance between $n_i(t, \bm x)\equiv G(t, \bm x| \bm x_i)$ as \cite{fp},
\begin{eqnarray}&&
H(t)=\int n_1(t, \bm x)\ln \frac{n_1(t, \bm x)}{n_2(t, \bm x)}d\bm x. \label{d1}
\end{eqnarray}
We observe that $H(t)$ is a non-negative function. To see that we recall that $\int n_i(\bm x, t) d\bm x=1$ and rewrite $H$ as
\begin{eqnarray}&&
H=\int \left[n_1\ln \frac{n_1}{n_2}-n_1+n_2\right]d\bm x
\nonumber\\&&
=\int n_2\left[
R\ln R-R+1\right]d\bm x,
\end{eqnarray}
where we introduced $R\equiv n_1/n_2$. The last term in the above equation is always non-negative as it follows
from
\begin{eqnarray}&&
R\ln R-R+1=\int_1^R \ln x dx\geq 0, \label{B6}
\end{eqnarray}
that holds for any $R\geq 0$. We therefore see that $H=0$ only at $n_1(t, \bm x)=n_2(t, \bm x)$. Thus $H(t)$ is non-negative and it vanishes only if $n_i, i=1,2$ are equal, thus providing a good definition of the distance between $n_i$.
Moving on, the rate of change of  $H$ with time is given by:
\begin{eqnarray}&&
{\dot H}=\int \left[\frac{n_1}{n_2}\nabla\cdot(\bm v n_2)-\ln \frac{n_1}{n_2}\nabla\cdot (\bm v n_1)
\right.\nonumber\\&&\left.
-\eta \frac{n_1}{n_2}\nabla^2 n_2+\eta \ln \frac{n_1}{n_2}\nabla^2 n_1 \right]d\bm x,
\end{eqnarray}
where we used the conservation constraint $\int n_1(\bm x, t) d\bm x=1$. Integrating now by parts yields:
\begin{eqnarray}&&\!\!\!\!\!\!\!\!
{\dot H}=
\int \left[n_1\bm v-n_1 \bm v \ln \frac{n_1}{n_2}
-\eta \frac{n_1}{n_2}\nabla n_2+\eta \ln \frac{n_1}{n_2}\nabla n_1 \right]d\bm S
\nonumber\\&&\!\!\!\!\!\!\!\!
-\eta \int n_1 \left(\nabla \ln (n_1/n_2)\right)^2 d\bm x,
\end{eqnarray}
where the surface integral is over the boundary of the domain of the flow. We assume that the boundary term can be discarded due to boundary conditions or because it is negligible in comparison with the volume term. We find,
\begin{eqnarray}&&\!\!\!\!\!\!\!\!
{\dot H}\approx -\eta \int n_1 \left(\nabla \ln (n_1/n_2)\right)^2 d\bm x\leq 0,
\end{eqnarray}
where equality holds for $n_1=n_2$ only. This equation implies relaxation to $n_1=n_2$ since $H\geq 0$, cf. Eq.~(\ref{so}). We conclude that $G(t, \bm x| \bm x')$ becomes independent of $\bm x'$ at $t \to\infty$ and thus:
\begin{equation}
B_z(t, \bm x)\approx G(t, \bm x| \bm x_0) \int  B_z(0, \bm x')d\bm x',
\label{Idependence}
\end{equation}
which leads to the following remarkable relationship:
\begin{equation}
B_z(t, \bm x)=I n_s(t, \bm x),\ \ I\equiv \int  B_z(0, \bm x)d\bm x 
\label{Idependence}
\end{equation}
where $\bm x_0$ is irrelevant. Here $n_s$ is defined in Eqs.~(\ref{sns})-(\ref{sno}), and $I$ denotes the (approximately) conserved spatial integral of $B_z$. For thin disks it is appropriate to call this the magnetic flux through the disk, disregarding difference in the definition that here is minor. Thus in our case of passive and active fields, there is no difference between them due to finite dissipation, $\eta>0$, see detailed discussion in \cite{fm}. We observe that the stationary measure is $z-$independent since in vertical dimension we have diffusion on a finite interval. Thus $B_z$ is $z-$independent in the steady state. 

We observe from Eq.~(\ref{Idependence}) that $B_z$ is sign-definite in the steady state. This is due to the decay of fluctuations as described by the case $I=0$. Moreover, with no loss, we may assume that $I\geq 0$, as can be guaranteed by the choice of the direction of $z-$axis, or by observing that we could use in the magnetohydrodynamic equations $-B_z$ instead of $B_z$ (indeed $-B_z$ and $B_z$ obey the same equation and generate the same Lorentz force). We conclude that the magnetic field endows the fluid with an effective density $|I|n_s(t, \bm x)$. The Lorentz force generated by the magnetic field resembles then the pressure force generated by this density.

We see that $B_z$ decays to zero if $I=0$ and is non-zero for $I\neq 0$. This can be understood by introducing the decomposition of $B_z$ into the sum of the constant mean (recalling that we use unit volume) and the fluctuation field $B'_z(t)$ that obeys,
\begin{eqnarray}&&\!\!\!\!\!\!\!\!\!\!\!\!\!\!
\partial_t B'_z\!+\!\nabla \!\cdot  \!( B'_z\bm v) \!=\!\eta\nabla^2  B'_z-I\nabla\cdot\bm v,
\end{eqnarray}
cf. Eq.~(\ref{generat}). If $I=0$ then $B'_z$ that has zero spatial integral decays due to mixing. The non-zero value of $I$ creates a stationary source of fluctuations and thus a non-trivial statistics of finite $B'_z$, cf. Eq.~(\ref{generat}). 

The above considerations, despite that they have a more complex form, are generalization of \cite{antidynamo}. They also hold for any planar velocity field, irrespective of whether the Lorentz force creates interaction of velocity and magnetic field. In the compressible case the (Lyapunov) functional of $\bm B$ that demonstrates convergence to a unique solution at large times, is significantly more complex than $\int B^2 d\bm x$ considered in \cite{antidynamo}, and is given by the "entropy" $H(t)$. However the approach is similar and so is that to the study of the horizontal components.  

{\bf Horizontal magnetic field}---We consider next the horizontal components of the magnetic field. As discussed above, under the current conditions the latter is derived from a scalar potential $A$ according to $B_x=\partial_y A$ and $B_y=-\partial_x A$. That potential obeys the advection-diffusion equation (\ref{adv}). That equation coincides with with its incompressible counterpart \cite{antidynamo} and, as
 in that case, it has constant solutions. However, to prove the long-time convergence to those solutions, a different approach must be devised. We use the properties of the Green's function to demonstrate that $A(t, \bm x)$ becomes constant at large times. Thus, we demonstrate that the Green's function ${\tilde G}(t, \bm x| \bm x')$ defined by,
\begin{eqnarray}&&\!\!\!\!\!\!\!\!\!\!\!\!\!\!
\partial_{t}{\tilde G}(t, \bm x| \bm x')\!+\!\bm v(t, \bm x)\!\cdot  \! \nabla {\tilde G}(t, \bm x| \bm x') \!=\!\eta\nabla^2 {\tilde G}(t, \bm x| \bm x'),\label{green1}
\end{eqnarray}
and ${\tilde G}(0, \bm x| \bm x')=\delta(\bm x-\bm x')$ becomes independent of $\bm x$ at $t\to\infty$. This implies that $A(t, \bm x)= \int {\tilde G}(t, \bm x| \bm x') A(0, \bm x')d\bm x'$ tends to a constant at large times. 

We consider the dependence of ${\tilde G}(t, \bm x| \bm x')$ on the time $t_0$ of setting the initial condition by defining $G(t, \bm x| t_0, \bm x')$ as the solution of Eq.~(\ref{green1}) that obeys ${\tilde G}(t_0, \bm x| t_0, \bm x')=\delta(\bm x-\bm x')$. As a function of the initial coordinate and time the Green's function obeys the Hermitian conjugate equation \cite{fp} so that,
\begin{eqnarray}&&\!\!\!\!\!\!\!\!\!\!\!\!\!\!
\partial_{t'}{\tilde G}'(t', \bm x'|\bm x)\!-\!\nabla'\!\cdot  \! \left( {\tilde G}'(t', \bm x'|\bm x)\bm v(t-t', \bm x')\right)
\nonumber\\&&\!\!\!\!\!\!\!\!\!\!\!\!\!\!
\!=\!\eta\nabla'^2 {\tilde G}'(t', \bm x'|\bm x),\ \ {\tilde G}'(t', \bm x'|\bm x)\equiv {\tilde G}(t, \bm x| t-t', \bm x'),
\end{eqnarray}
where ${\tilde G}'(t'=0, \bm x'|\bm x)=\delta(\bm x-x')$ and we designated derivative with respect to $\bm x'$ by prime. We observe that ${\tilde G}'(t'=0, \bm x'|\bm x)$ obeys Eq.~(\ref{green}) at large times and thus becomes independent of $\bm x$ at large times as demonstrated previously. This implies independence of ${\tilde G}(t, \bm x| \bm x')$ of $\bm x$ at $t\to\infty$. Thus $A$ becomes constant at large times which by taking derivatives implies $B_x=B_y=0$.

{\bf Summary}---We conclude from the above that, provided that the contribution of the boundary contributions is negligible on the considered times, at large times the magnetic field becomes a stationary vertical field. It is proportional to the stationary measure $n_s(t, \bm x)$ which has a unit space integral, with proportionality coefficient given by the constant magnetic flux. That measure provides long time behavior of the solutions to Eq.~(\ref{sns}). Thus, there is no indefinite growth of the field, such as that occurring in three dimensions (the growth can occur in a transient manner, cf. \cite{igor}) but rather settling into a steady state field. 

{\bf Zeldovich's theorem generalizes in the zero flux case}---If the flux vanishes, $I=0$, 
then the above implies that the magnetic field relaxes to zero. Thus we can formulate the statement: planar flows with zero magnetic flux do not support magnetic fields in the steady state, irrespective of compressibility, cf. \cite{antidynamo,ll8}. 

{\bf Finite flux and effective equations of the magnetohydrodynamics}---In the case of $I\neq 0$ (recall that $I$ represents the magnetic flux that is associated with the $z$-direction, see Eq. (\ref{Idependence})), 
%AK: Shall we remind here that the non-zero flux is associated with z-direction?
the magnetic field is non-trivial in the steady state. The momentum equation (\ref{lorne}) takes the form
\begin{eqnarray}&&\!\!\!\!\!\!\!\!\!\!\!\!
\partial_t\bm v+(\bm v\cdot\nabla)\bm v=-\frac{\nabla \ln \rho}{Ma^2}+\bm a+\nu\nabla^2\bm v%\nonumber\\&&
-\frac{I^2}{2\rho}\nabla n_s^2, \label{redud}
\end{eqnarray}
where $n_s$ is the steady state solution of Eq.~(\ref{sns}). Thus the action of the magnetic field on the velocity boils down to the magnetic pressure which is similar to ordinary pressure in a gas with a polytropic index $2$. The role of the Mach number is played by $I^{-1}$, where the role of sound velocity is played by the (group) velocity of Alfv\'en waves \cite{ll8}.

{\bf Role of the boundary conditions}---The above conclusions rely on boundary conditions that probably are not realistic. For instance for the spatial integral of $B_z$ (the magnetic flux) we have from Eq.~(\ref{sgo}) that 
\begin{eqnarray}&&\!\!\!\!\!\!\!\!\!\!\!\!
\dot I=\frac{d}{dt} \int B_zd\bm x\!=\!\eta\int \hat {\bm n}\cdot\nabla  B_z ds, \label{nc}
\end{eqnarray}
where the last integral is over the disk boundary which normal is designated by ${\hat n}$. The other boundary integral appearing in the calculations, $\int ds B_z \hat {\bm n}\cdot\bm v$, is assumed to vanish. This is true if there is no flux of matter through the boundary ($\hat {\bm n}\cdot\bm v=0$) or $B_z$ vanishes at the boundary as it was assumed in \cite{antidynamo}. We see that the rate of change of the magnetic flux, which by itself is proportional to the area, is proportional to the disk's perimeter. Thus in the large area limit the non-conservation of the flux is negligible. Similar conclusions hold for other boundary effects. In other words, the conclusions, obtained using artificial finite box conditions, hold over a finite time interval, that becomes infinite in the thermodynamic limit. In applications, this signifies that adiabatic approach must be used where $I$ is considered as a variable that slowly varies in time, and possibly also space. 

{\bf Conditions for neglecting the magnetic field in steady state}---We saw that at $I=0$ there is no magnetic field in the steady state, whereas in the limit of strong field, $I\to\infty$, the magnetic field generates Lorentz force that influences the flow. In between these opposite limits there is a critical amplitude of the flux $I_{cr}$, defined by the condition that we can neglect the magnetic field at $I\ll I_{cr}$, but cannot otherwise. 

In order to determine $I_{cr}$ we use self-consistency argument. We first assume that the 
magnetic field is negligible. Then it is a passive field so that the flow is purely hydrodynamic, as considered previously. Recalling that $n_s$ is roughly $\rho$ coarse-grained over the scale $l_{\rm res}$, we can employ the insight into $\rho$ gained from the analysis of pure hydrodynamic compressible turbulence laid out in the previous sections, in order to estimate $n_s=B_z/I$. Finally we plug the obtained $n_s$ into Eq.~(\ref{redud}), evaluate the resulting Lorentz force and check the assumption that it is negligible. The value of $I_{cr}$, obtained in this way depends on the Mach number and is considered below. 

{\bf Magnetic field at $Ma\ll 1$}---The considerations here resemble those of the strong field limit above, where the role of external field is played by the average magnetic field $I$. Under the assumption that magnetic field is negligible, in pseudo-sound regime, the density modulation is of order $Ma^2$, i.e. $\rho=1+O(Ma^2)$, see e.g. \cite{levermore,pseudos}. It seems inevitable that in all cases of practical interest we have $n_s=1+\delta n$ with $\delta n= \sigma Ma^{\beta}+o(Ma)$, where the actual value of the positive exponent $\beta$, that most probably is equal to $2$, is irrelevant below. Here the field $\sigma$ is independent of $Ma$ and is of order one, unless there is a significant numerical factor, that is beyond the parameteric estimates performed here. We find $B_z=I+I\delta n$ and Eq.~(\ref{redud}) becomes 
\begin{eqnarray}&&\!\!\!\!\!\!\!\!\!\!\!\!
\partial_t\bm v\!+\!(\bm v\!\cdot\!\nabla)\bm v\!=\!-\frac{\nabla \ln \rho}{Ma^2}\!+\!\bm a\!+\!\nu\nabla^2\bm v
%\nonumber\\&&
-I^2 Ma^{\beta}\nabla \sigma, \label{mp}
\end{eqnarray} where the magnetic pressure is written to the leading order. We see that the magnetic pressure term (Lorentz force) is not negligible if $I Ma^{\beta/2}\gtrsim 1$. For instance, for $\beta=2$ we find that the magnetic field is relevant if $I\gtrsim 1/Ma$, cf. the discussion of external magnetic field above. In other words, however small $Ma$ is, for large enough $I$, the compressibility changes significantly the flow which acquires the magnetic degree of freedom that interacts appreciably with the velocity. Moreover, if $I\gg Ma^{-\beta}$ then the magnetic pressure makes a large change in the flow - the assumption of order one change would not be consistent with Eq.~(\ref{mp}). 

A remark similar to that made in the discussion of behavior in external field must be made. The low Mach number density field was not studied much in the two-dimensional case. The inverse cascade demands care in considering the magnitude of $\nabla\sigma$ at different scales and its relevance at a given scale. We leave this for future work. The above considerations seem to make it inevitable that in the low Mach number limit $I_{cr}\propto Ma^{-\gamma}$ with $\gamma>0$.

The above considerations reaffirm that care is necessary in using the Zeldovich antidynamo theorem. This theorem, that could be hoped to apply at $Ma\ll 1$, fails if there is a large magnetic field threading the disk.   

{\bf Magnetic field at $Ma\sim 1$}---In the case of $Ma\sim 1$, the density is a smooth large-scale field with order one fluctuations, see Fig.~\ref{Fig:M096}. The structure of $n_s=B_z/I$ is similar - the difference of passive and active fields in this case is less relevant \cite{fm}. We see from Eq.~(\ref{redud}) that for $I\ll 1$ the magnetic field is negligible and for $I\sim 1$ it has order of one effect on the flow. For $I\gg 1$ the magnetic field leads to significant rearrangement of the velocity field. In other words, $I_{cr}\sim 1$. We assumed that $l_{\rm res}\lesssim L$, the case which seems most practical. 

%In the range of small Mach numbers two-dimensional turbulence manifests inverse cascade of energy \cite{fk} familiar from the incompressible case holding at zero Mach number, see, e.g., \cite{review}. The largest scale of the turbulent eddies is given either by the size of the system or the sonic scale $l_s$ at which the velocity of the turbulent eddies is comparable with the speed of sound \cite{fk,fmsonic}. Furthermore, in the small Mach number regime the compressibility of the flow is not large so that
{\bf Magnetic field at $Ma\gg 1$}---Finally, we turn to the case of large Mach numbers. In this regime the reversal of the energy cascade direction discussed in the sections above comes into force and plays a most significant role in the magnetic field amplification process. Indeed, as discussed above, the direction of the cascade is now almost fully direct, just as in the three dimensional case. As a result, the density field that is generated by the continuity equation is multifractal, in a stark contrast to the smooth density fields in the $Ma\lesssim 1$ case. Furthermore, at the large Mach numberss regime the density field is passive, see Eq.~(\ref{ns}). Therefore, making our probe assumption that the impact of the magnetic field on the velocity is negligible, $B_z/I$ at $\eta=0$ and $\rho$ coincide. Thus, we have from Eq.~(\ref{fls}) that at $\eta\to 0$ (here $\left\langle B^2\right\rangle\approx \left\langle B_z^2\right\rangle$)
\begin{eqnarray}&&\!\!\!\!\!\!\!\!\!\!\!\!\!\!
\left\langle B^2\right\rangle \simeq I^2 C_0C^{\delta}Ma^{2\delta},\ \ \frac{\left\langle B^2\right\rangle }{\left\langle B_z\right\rangle^2} \simeq C_0C^{\delta}Ma^{2\delta}. \label{fs}
\end{eqnarray}
For other aspects of (multifractal) statistics of $B_z$ at $\eta=0$, see \cite{fm}. The smoothness scale of $B_z$ at $\eta=0$ is the sonic scale $l_s$, which we assume here to obey $l_s\gtrsim l_{\rm res}$. Then Eq.~(\ref{fs}) holds also at finite $\eta$. We find that magnetic energy is increased by a power of $Ma$, as compared to the mean field estimate. 

Magnetic energy however is determined by the sonic scale and it cannot serve to decide whether the magnetic field is relevant at large scales. For studying these scales, we assume the condition $l_{\rm res}\ll l_s$, which implies $n_s=\rho$. Then the reduced Eq.~(\ref{redud}) becomes at scales larger than $l_s$ 
\begin{eqnarray}&&\!\!\!\!\!\!\!\!\!\!\!\!
\partial_t\bm v+(\bm v\cdot\nabla)\bm v=\bm a+\nu\nabla^2\bm v
-I^2\nabla \rho.\label{red}
\end{eqnarray} 
The equations have the form of the shallow water model. 
The relevance of magnetic pressure at large scales is determined by $I$, as in $Ma\sim 1$ case. This is because at large scales the density is of order one and has fluctuations of order one, as at $Ma\sim 1$. Thus if $I\ll 1$ the magnetic field is irrelevant at the injection scale, for $I\sim 1$ it changes the large-scale flow, and at $I\gg 1$ the magnetic field brings a large change in the flow at scale of the pumping. 

If $I\ll 1$, the magnetic pressure term is irrrelevant at large scales. However, as in the previous discussion of ordinary thermodynamic pressure, it becomes relevant at smaller scales. Thus the last term in Eq.~(\ref{red}) is comparable with the advective term at the ``magnetosonic scale'' $l_m$ where eddy velocity and (group) velocity of Alfv\'en waves are of the same order. Thus, at  $l_m$ the squared eddy velocity is of order of $I^2$ times the density. Using for estimates $k^{-2}$ spectrum for the velocity, and $(L/l_m)^{\delta/2}$ for density at the scale $l_m$, see Eq.~(\ref{fls}) and \cite{fm}, we find 
$l_m\sim I^{2/(1+\delta)} L$. By including the thermal pressure, considered previously, we conclude  that if $I^{2/(1+\delta)}>1/Ma^2$, then the cascade is stopped by magnetic pressure at the scale $l_m$, and otherwise it is stopped by the thermal pressure at $l_s$.

We comment on the assumption that $n_s=\rho$ under the condition $l_{res}\ll l_s$. It could be that the limits of vanishing $\eta$ and infinite evolution time do not commute per realization. Then $n_s=\rho$ is only an equality in law and not an ordinary equality. This would make Eqs.~(\ref{red}) valid qualitatively only. This fits our purposes here and we leave more detailed study for future work.   

{\bf Time of validity of neglecting the details of the boundary conditions}---The time of formation of the level of fluctuations described by Eq.~(\ref{fs}) is determined by the typical time for the cascade process to reach $l_s$ when starting from $L$, see \cite{fm}. This time is of order $L/u$ and is independent of $l_s$, as in incompressible turbulence \cite{frisch}. For instance, if we start with uniform initial $B_z$ (which is the simplest case of $I\neq 0$), then Eq.~(\ref{fs}) is valid at times larger than $L/u$ and holds until times where the boundary effects become relevant (this is true as long as the reaction of the magnetic field on the flow is negligible). Thus our treatment of the boundary conditions is self-consistent for the energy calculation, provided that the characteristic time at which the boundary phenomena penetrate the bulk of the disk is much larger than $L/u$.  

{\bf Numerical calculation of the Green's function}---We conclude this section by remarking on the numerical calculation of the evolution of $B_z$ as described by $G(t, \bm x| \bm x')$. The Green's function can be found as the PDF of $\bm x(t)$ that is obtained by solving the stochastic ODE
$\dot{\bm x}=\bm v(t, \bm x(t))+\sqrt{\eta}\bm \Gamma$ with the initial condition $\bm x(0)=\bm x'$. Here $\bm \Gamma$ is a white noise with the dispersion $\langle \Gamma_i(t)\Gamma_k(t')\rangle=2\delta_{ik}\delta(t-t')$, see details in \cite{fp}. Solving many realizations of the ODE necessary for the statistics is simpler than solving the PDE. Alternatively, the solution is given by the Feynman-Kac formula \cite{fp},
\begin{eqnarray}&&\!\!\!\!\!\!\!\!\!\!\!\!\!\!
G(t, \bm x| \bm x')=\int_{\bm x(0)=\bm x',\ \ \bm x(t)=\bm x} {\cal D}\bm x
\nonumber \\&&\!\!\!\!\!\!\!\!\!\!\!\!\!\!
\exp\left(-\frac{1}{4\eta}\int_0^t \left(\dot {\bm x}-\bm u(t, \bm x(t'))\right)^2dt'\right), \label{green}
\end{eqnarray}
see \cite{fp}. The solution is valid also when the velocity field interacts with the magnetic field so it can be used for simulations of the full system of magnetohydrodynamic equations (MHD). Eq.~(\ref{green}) allows to see that at large times $G(t, \bm x| \bm x')$ as a function of $\bm x$ is independent of $z$, assuming that $z$ belong to a fixed finite interval. Thus also the magnetic field is independent of $z$ as already said. This is readily seen from the decoupling $\left(\dot {\bm x}-\bm u(t, \bm x(t'))\right)^2=\sum_{k=1}^2 \left(\dot x_k-u_k(t, \bm x(t'))\right)^2+{\dot z}^2$ in the weight function in Eq.~(\ref{green}).

{\bf Burgers' model for magnetohydrodynamic accretion at large $Ma$ and $I\ll 1$}---We consider now the implications of the above for the structure of solutions of the full system of magnetohydrodynamic equations. We focus our attention on the $Ma\gg 1$ case. We assume that $I\ll 1$ so that the magnetic field is negligible at the pumping scale. We demonstrated previously that the cascade is stopped by magnetic pressure at the scale $l_m$, and otherwise it is stopped by the thermal pressure at $l_s$. Taking these considerations into account, the following equations provide a generalisation of the Burgers' model for accretion that is given by Eq.~(\ref{proposed}), by including the effects of magnetic fields:  
\begin{eqnarray}&&\!\!\!\!\!\!\!\!\!\!\!\!\!\!
\partial_t\bm v+(\bm v\cdot\nabla)\bm v=\bm a+\left(\frac{b}{Ma^3}+b'I^{3/(1+\delta)}\right)\nabla^2\bm v,
\end{eqnarray}
where $b'$ is a constant of order one and the cofficient of the Laplacian term guarantees that the smoothness scale of the flow is of order $max[l_m, l_s]$. The density and magnetic field then obey the corresponding equations 
\begin{eqnarray}&&\!\!\!\!\!\!\!\!\!\!\!\!\!\!
\partial_t \rho+\nabla\cdot(\rho\bm v)=\left(\frac{b}{Ma^3}+b'I^{3/(1+\delta)}\right)\nabla^2\rho,\nonumber\\&&\!\!\!\!\!\!\!\!\!\!\!\!\!\!
\partial_t B_z\!+\!\nabla \!\cdot  \!( B_z\bm v) \!=\!\left(\frac{b}{Ma^3}+b'I^{3/(1+\delta)}\right)\nabla^2  B_z,
\end{eqnarray}
%AK: is B a scalar here? -- sorry, I got lost:)
with reasoning similar to that of Eq.~(\ref{proposa}). We do not consider the case where the resistive scale is larger than $l_s$ since it seems of little practical interest and can be treated readily. The advantage of the above model is similar to that of the Burgers' model for accretion without magnetic field, considered previously. We leave for future work applications of the above framework. 

{\bf Conclusions}---As the Mach number of the flow in thin disks increases, the sonic scale $l_s$ decreases from a value much larger than the integral scale $L$, for $Ma\ll 1$, to a value of order $L$ at $Ma\sim 1$, and, finally, to a value much smaller than $L$ at $Ma\gg 1$. This decrease of $l_s/L$ is accompanied by the reversal of the direction of energy cascade from inverse in the case of small Mach numbers to direct cascade for supersonic turbulence. We can use the ratio of the kinetic energy fluxes to large and small scales, cf. \cite{fk} as an order parameter for this transition.  That ratio changes continuously as a function of the Mach number so that the transition is smooth.  

Supersonic turbulence occurs at $Ma\gg 1$ in an interval of scales between $l_s$ and $L$, the supersonic inertial range. This range widens indefinitely as $Ma\to\infty$ and is characterised by multifractal fluctuations of the density inside it, just as in supersonic three-dimensional turbulence \cite{dust,fm}. 

In order to gain deeper insight into compressible hydrodynamic turbulence we have derived an explicit expression for the third order velocity correlation tensor in the case of potential driving force. A significant result of that derivation is that to lowest order all third order structure functions are zero. As discussed above this reflects the nature of the physical processes that take place in highly compressible turbulence and may have significant implications for observations as well as numerical simulations.

The possibility of the emergence of magnetic fields in compressible turbulent conducting fluids has been considered. Remarkably, we demonstrate that the limit of incompressible flow, $Ma\to 0$, has singular features in the presence of mean vertical magnetic field (flux) $I$. The limits of $Ma\to 0$ and $I\to\infty$ do not commute. Thus for however small $Ma$, fields with $I\gtrsim Ma^{-\gamma}$ cause the flow to differ appreciably from incompressible purely hydrodynamic flows, that hold at $Ma=0$ according to the antidynamo theorem. Incompressible flow is an idealization that is valid only if $I\ll Ma^{-\gamma}$.   

We performed a systematic
%AK "performed a systematic study of ..."
study of the role played by the magnetic field in 
%AK: "...play in different Mach number regimes."
different Mach number regimes. We provided a generalisation of the antidynamo theorem to compressible flows. We demonstrated that the horizontal components of initially small magnetic field fluctuations decay in time, as in the incompressible case. The antidynamo theorem generalizes fully at $I=0$, 
%AK: "where the fluctuations in the vertical component decay as well."
where the vertical component decays also. The result of decaying magnetic fluctuations in the zero mean field is compatible with 
%AK "...is supported by..."
numerical simulations of magnetohydrodynamic flows in astrophysical disks (see e.g. \cite{cattaneo} and \cite{pessah}).

The situations with finite flux can be understood by introducing a $Ma-$dependent threshold $I_{cr}$. The case of $I\ll I_{cr}$ is quite similar to $I=0$. Here the magnetic field is stationary and non-trivial. However, its steady state fluctuations generate weak Lorentz forces so the magnetic field does not change velocity appreciably. The fluctuations are significantly amplified at $Ma\gg 1$ where the magnetic field energy is enhanced significantly in comparison with the square of the mean field values,
%AK: "... mean field values,"
due to the multifractal nature of the density. Care is needed since magnetic field can be negligible at some scales and relevant at others. 

In the limit of $I\gtrsim I_{cr}$ the Lorentz forces are appreciable and we have a steady state characterized by strong coupling of magnetic and hydrodynamic degrees of freedom. This state is much simpler, though than in three dimensions. The Lorentz force reduces to magnetic pressure forces which are similar to pressure in the shallow water model. The resemblance is reinforced by the observation that with no loss the steady state magnetic field is equivalent to a positive density. 

Our results can be used to provide predictions for observations of thin astrophysical disks. A robust implication of our study is that for planar flow horizontal magnetic field is zero in the steady state. Thus, observations of horizontal components in thin astrophysical disks might be highly informative. Their presence would signify that two-dimensional modelling is inaccurate and there is some $z-$dependence of $B_z$. 

Another property of much observational interest is the ratio $B_z/\rho$. We recall that in ideally conducting fluid the ratio of magnetic field and the density is simply transported along the Lagrangian (fluid particle) trajectories of the fluid \cite{ll8}. Here it might have simpler properties, say at $Ma\gg 1$ both $B_z$ and $\rho$ are multifractal yet their ratio must be much smoother (if not a constant). 

Finally, we proposed Burgers' model of accretion as a way to provide effective
%AK: robust?
description of the flow at large $Ma$. Since in the case of gravity driven turbulence the flow in two dimensions is potential, then the model is equivalent to a linear equation, after a Cole-Hopf transformation. This is a very promising direction for numerical studies since the solutions to the equations are then much more stable and moreover can be obtained via path integral solution. Considerable interaction between theory and observations is necessary to ensure steady progress. 

This work was supported by the BSF grant no. 8773312. The simulations utilized a resource allocation MCA07S014 of the ACCESS program, which is supported by NSF Grants \#2138259, \#2138286, \#2138307, \#2137603, and \#2138296. A.K. was supported in part by the NASA Grant 80NSSC22K0724.

%We consider the generalization of the above considerations to the axially symmetric case. The magbne
%%%%%%%%%%%%%%%%%%%%%%%%%%%%%%%%%%%%%%%%%%%%%

%%%%%%%%%%%%%%%%%%%%%%%%%%%%%%%%%%%%%%%%%%%%%

\end{document}